\documentclass[12pt]{article}
\usepackage{amssymb,amsmath,amscd}
\usepackage{epsfig}
\usepackage{epstopdf}
\usepackage{latexsym}
\usepackage{graphicx}
\usepackage{subfigure}
\usepackage{booktabs}
\usepackage{bbm}
\usepackage[margin=20pt,small]{caption}
\usepackage[vcentermath,stdtext]{youngtab}
\usepackage[toc]{appendix}
\usepackage[numbers,compress]{natbib}
\usepackage{natbib}
\usepackage{color}
\usepackage{datetime}
\usepackage[
      colorlinks=false,
      linkcolor=darkblue,  
      urlcolor=blue,    
      filecolor=blue,     
      citecolor=red,
linktocpage=true,
      pdfstartview=FitV,
      bookmarksopen=true    
      ]{hyperref}

\ifpdf
\fi

\renewcommand{\theequation}{\arabic{section}.\arabic{equation}}

\def\coeff#1#2{\relax{\textstyle {#1 \over #2}}\displaystyle}

\def\ZZ{\mathbb{Z}}

\def\cN{{\cal N}}

\def\cR{{\cal R}}

\def\eql{=}
\def\CC{\mathbb{C}}
\def\RR{\mathbb{R}}
\def\PP{\mathbb{P}}

\definecolor{cardinal}{rgb}{0.6,0,0}
\definecolor{darkgreen}{rgb}{0,0.5,0}
\definecolor{golden}{rgb}{0.92, 0.7, 0}
\definecolor{midnight}{rgb}{0, 0, 0.5}
\definecolor{darkblue}{rgb}{0.2, 0, 0.8}
\newcommand{\Red}{\color{red}}

\def\bfs#1{{\boldsymbol #1}}

\def\cals#1{\mathcal{#1}}

\def\eo{\overset{_{\phantom{.}\circ}}{e}{}}

\def\go{\overset{_{\phantom{.}\circ}}{g}{}}
\def\Do{\overset{_{\phantom{.}\circ}}{D}{}}

\def\Ro{\overset{_{\phantom{.}\circ}}{R}{}}

\def\SE{{\sc se}}
\def\FR{{\sc fr}}
\def\PW{{\sc pw}}
\def\KE{{\sc ke}}
\def\KK{{\sc kk}}
\def\BF{{\sc bf}}

\def\stsx{\mathop{\bullet}}
\def\stsv{\mathop{\ast}}

\def\dGamma{{\rm I\!\Gamma}}

\def\Ga{\Gamma}

\def\??{{\Red \bf ??}}

\begin{document}  

\begin{titlepage}
\bigskip

\bigskip
\bigskip
\begin{center}
{\Large \bf
On perturbative instability of Pope-Warner solutions \\[6 pt] on  Sasaki-Einstein manifolds }

\bigskip\bigskip\bigskip\bigskip

{\bf Krzysztof Pilch and Isaiah Yoo \\ }
\bigskip
Department of Physics and Astronomy \\
University of Southern California \\
Los Angeles, CA 90089, USA  \\
\bigskip
pilch@usc.edu, isaiahyo@usc.edu  \\
\end{center}

\vspace{1cm}
\begin{abstract}
{
\small

Given a   Sasaki-Einstein manifold, $M_7$,   there is  the  $\cN\!=\!2$ supersymmetric $AdS_4\times M_7$ Freund-Rubin solution of eleven-dimensional supergravity and the corresponding non-supersymmetric solutions: the perturbatively stable skew-whiffed solution, the perturbatively unstable  Englert solution,  and the  Pope-Warner solution, which is known to be perturbatively unstable when  $M_7$ is  the seven-sphere or, more generally, a tri-Sasakian manifold.   We show that similar  perturbative instability of the Pope-Warner solution will arise for any Sasaki-Einstein manifold, $M_7$, admitting a basic, primitive, transverse (1,1)-eigenform of the   Hodge-de Rham Laplacian with the eigenvalue in the range between $2(9-4\sqrt 3)$ and $2(9+4\sqrt 3)$. Existence of such  (1,1)-forms on all homogeneous Sasaki-Einstein manifolds can be  shown explicitly using the K\"ahler quotient construction or the standard harmonic expansion. The latter shows that the instability arises from the coupling between the Pope-Warner background  and   Kaluza-Klein scalar modes that at the supersymmetric point lie in a long $Z$-vector supermultiplet.  
We also   verify that the instability  persists for the orbifolds  of homogeneous Sasaki-Einstein manifolds that have been discussed recently.  
}
\end{abstract}
\end{titlepage}


\tableofcontents

\section{Introduction}
\label{Intro}

The AdS/CFT correspondence \cite{Aharony:1999ti} and   ABJM theory \cite{Aharony:2008ug}  have led to renewed interest in solutions  of eleven-dimensional supergravity \cite{Cremmer:1979up} whose equations of motion in the bosonic sector are:\footnote{We summarize our conventions in appendix \ref{appendixA}.}
\begin{equation}\label{Mth:eineqs}
\cals R_{MN}+  \frak g_{MN}\cals R\eql {1\over 3} \cals F_{MPQR}\cals F_{N}{}^{PQR}\,,
\end{equation}
\begin{equation}\label{Mth:maxeqs}
d\star \cals F_{(4)}+\cals F_{(4)}\wedge \cals F_{(4)}=0\,,
\end{equation}
where $\frak g_{MN}$ is the metric,  $\cals F_{(4)}=d\,\cals A_{(3)}$ is the four form flux, and  $\star$ denotes the Hodge dual in eleven dimensions. A simple and important class of solutions  are the ones in which the eleven-dimensional space time is a product $AdS_4\times M_7$, where   $M_7$  is a seven-dimensional Sasaki-Einstein (\SE) manifold. 
Those  manifolds are  characterized by the existence of two real Killing spinors (see, e.g., \cite{Boyer:2007nr,Galicki:book,Sparks:2010sn}) and the corresponding Freund-Rubin (\FR) solutions \cite{Freund:1980xh} are $\cals N\geq 2$ supersymmetric.
Solutions in  which $M_7$ is one of the homogeneous \SE\ manifolds:%
\begin{equation}\label{homSE}
S^7\,,\qquad N^{1,1}\,,\qquad M^{3,2}\,,\qquad Q^{1,1,1}\,,\qquad V^{5,2}\,,
\end{equation}
were classified in the  1980s  \cite{Castellani:1983yg}, but it is only quite recently that new solutions with nonhomogeneous  \SE\ metrics have been discovered \cite{Gauntlett:2004hh,Cvetic:2005ft,Cvetic:2005vk}. 

It has also been known since the 1980s that given a SE manifold, $M_7$, there are, in addition to the supersymmetric \FR\ solution, three non-supersymmetric solutions: the skew-whiffed \FR\ solution \cite{Freund:1980xh} obtained by the change of orientation on $M_7$, and  the Englert  \cite{Englert:1982vs} and Pope-Warner (\PW) \cite{Pope:1984bd,Pope:1984jj} solutions with nonvanishing  internal fluxes constructed from the  geometric data on $M_7$. Those solutions have recently turned out in ``top-down'' constructions of holographic superconductors \cite{Gauntlett:2009dn,Gubser:2009gp,Gauntlett:2009bh}, where they correspond 
to  critical points of the scalar potential in various truncations \cite{Warner:1983vz,Gauntlett:2009zw,Bobev:2010ib,Cassani:2011fu,Cassani:2012pj} of eleven-dimensional supergravity. 

Quite generally, non-supersymmetric solutions in gauged supergravity tend  to be unstable. Indeed, while the stability of an $AdS$-type solution is guaranteed if there are some unbroken supersymmetries \cite{Abbott:1981ff,Gibbons:1983aq}, in  non-supersymmetric backgrounds one expects to find scalar fluctuations,
\begin{equation}\label{sceqs}
(\Box_{AdS_4}-m^2)\varphi \eql 0\,,
\end{equation}
whose masses  violate the Breitenlohner-Freedman (\BF) bound \cite{Breitenlohner:1982jf}
\begin{equation}\label{BFbound}
m^2L^2\geq - {9\over 4}\,,
\end{equation}
where $L$ is the radius of $AdS_4$.

For the solutions above, the perturbative stability of the skew-whiffed \FR\ solution in eleven-dimensional supergravity was proved in \cite{Duff:1984sv}. It follows by a simple observation that the  mass spectrum of  fluctuations that might produce an instability is invariant under the change of orientation of $M_7$ and hence is the same for the skew-whiffed and the supersymmetric backgrounds. 

The Englert solutions are more difficult to analyze because the background flux couples the scalar and  pseudoscalar fluctuations. The resulting   perturbative instability  for any \SE\ background, $M_7$,  can be shown by an explicit construction of unstable modes in terms of the two Killing spinors  \cite{Page:1984fu}.  The same instability is also visible in the massive truncation of the eleven-dimensional supergravity on $M_7$ \cite{Gauntlett:2009zw,Gauntlett:2009bh}, and  when $M_7$ is the round seven-sphere, $S^7$, it corresponds to  the instability of the $\rm SO(7)^-$ critical point of $\cals N=8$, $d=4$ gauged supergravity~\cite{deWit:1983gs,Biran:1984jr}.

The  question of stability of the \PW\ solutions is more subtle as those turn  out to be stable within the massive  truncation  of \cite{Gauntlett:2009bh,Gauntlett:2009zw}.  However, in the special case of $S^7$, it  was  shown that the minimal sector of this truncation coincides with the $\rm SU(4)^-$ invariant sector of $\cals N=8$, $d=4$ gauged supergravity, within which the critical point of the potential corresponding to the \PW\ solution was found to be unstable \cite{Bobev:2010ib}. The unstable modes in the $\cals N=8$, $d=4$ theory transform in $\bf 20'$ of  $\rm SU(4)^-$  and can be lifted to the eleven-dimensional supergravity  \cite{Bobev:2010ib}, where the $\rm SU(4)$ symmetry becomes the isometry of $\CC\PP^3$ -- the Kahler-Einstein base of the Hopf fibration of $S^7$. 

Subsequently, the \PW\ solutions have been shown to be unstable within a massive truncation on any tri-Sasakian manifold \cite{Cassani:2011fu}. In this construction, a single unstable mode crucially depends on an additional Killing spinor beyond  the canonical two defining the particular \SE\ structure of the \PW\  solution.  From this perspective, the twenty unstable modes on $S^7$ can be understood as the symmetric, traceless product  of $\bf 6$ of $\rm SU(4)$ corresponding to six additional Killing spinors on $S^7$. 

A generalization of the truncation in \cite{Gauntlett:2009bh,Gauntlett:2009zw}, which included additional modes on homogeneous Sasaki-Einstein manifolds, was constructed in \cite{Cassani:2012pj},  and it was found that within that truncation the \PW\ solution remained stable. One might then naively conclude that the instabilities above should be attributed to additional Killing spinors and hence be absent for a sufficiently generic \SE\ manifold.

In this paper we identify a potential source of perturbative instability of the \PW\ solution on an arbitrary (regular) \SE\ manifold. We show that starting with a basic, primitive, transverse (1,1)-form\footnote{In the following, also referred to as the ``master form.''} $\omega$ on $M_7$, which is an eigenform of the  Hodge-de Rham Laplacian, $\Delta_2$, with the eigenvalue $\lambda_{\omega}\geq 0$, one can construct explicitly one metric and two flux harmonics, which after diagonalization of the linearized equations of motion give rise to three modes in the scalar spectrum with the following masses:
\begin{itemize}
\item [(i)] supersymmetric \FR\ 
\begin{equation}\label{rFRmass}
m^2L^2\,:\qquad {\lambda_\omega\over 4}-2\,,\qquad {\lambda_\omega\over 4}+\sqrt{\lambda_\omega+1}-1\,,\qquad {\lambda_\omega\over 4}-\sqrt{\lambda_\omega+1}-1\,,
\end{equation}
\item [(ii)] skew-whiffed \FR\  
\begin{equation}\label{swFRmass}
m^2L^2\,:\qquad  {\lambda_\omega\over 4}-2\,,\qquad
{\lambda_\omega\over 4}+2\sqrt{\lambda_\omega+1}+2\,,\qquad 
{\lambda_\omega\over 4}-2\sqrt{\lambda_\omega+1}+2\,,
\end{equation}
\item [(iii)] \PW\
\begin{equation}\label{PWmass}
m^2L^2\,:\qquad {3\over 8}\,\lambda_\omega\,,\qquad {3\over 8}\,\lambda_\omega+3\sqrt{\lambda_\omega+1}+3\,,\qquad {3\over 8}\,\lambda_\omega-3\sqrt{\lambda_\omega+1}+3\,.
\end{equation}
\end{itemize}

For the first two solutions, all modes in \eqref{rFRmass} and \eqref{swFRmass} are stable with the lowest possible masses saturating the BF-bound  \eqref{BFbound} when $\lambda_\omega=3$ and $\lambda_\omega=15$, respectively. However, for the \PW\ solution, the last mode in \eqref{PWmass} becomes unstable when $\lambda_\omega$ lies in the range
\begin{equation}\label{PWrange}
2(9-4\sqrt 3)<\lambda_\omega< 2(9+4\sqrt 3)\,.
\end{equation}
In principle, all that remains then is to determine which  \SE\ manifolds admit such stability violating $(1,1)$-forms.  Unfortunately, this appears to be a difficult problem since no general bounds on the low lying eigenvalues of  $\Delta_2$  on an arbitrary \SE\ manifold are known.

\begin{table}[t]
\begin{center}
{
\begin{tabular}{@{\extracolsep{20 pt}} c c c c c c}
\toprule
\noalign{\smallskip}
\quad$M_7$ & $\lambda_\omega$ & $m^2L^2$ &   $\#$ of modes & \KK\ spectra  \\
\noalign{\smallskip}
\midrule
\noalign{\smallskip}
$S^7$ & 24 & $-3$ &   20 & \cite{cp3eigenvalues,Duff:1986hr}\\[6 pt]
$N^{1,1}$ & 24 & $-3$ &   1 &  \cite{Termonia:1999cs,Fre':1999xp,Billo:2000zr} \\[6 pt]
$M^{3,2}$ & 16 & $9-3\sqrt{17}$  &   8 & \cite{D'Auria:1984vv,Castellani:1991et,Fabbri:1999mk} \\[6 pt]
$Q^{1,1,1}$ & 16 & $9-3\sqrt{17}$ &     9 & \cite{Merlatti:2000ed} \\[6 pt]
$V^{5,2}$  & ${32/ 3}$ & $7-\sqrt{105}$  &  5 &   \cite{Ceresole:1999zg} \\[6 pt]
\bottomrule
\end{tabular}
}
\caption{\label{tabone} Unstable modes for the \PW\ solution on homogeneous \SE\ manifolds.
{ 
}}
\end{center}
\end{table}

In the absence of general results, we look at the homogeneous  \SE\ manifolds \eqref{homSE} for which the spectra of the Hodge-de Rham Laplacians, $\Delta_k$, and of the Lichnerowicz operator, $\Delta_L$, have been calculated in the references listed in Table~\ref{tabone}, either as part of the  Kaluza-Klein program in the 1980s,\footnote{For a review, see, e.g., \cite{Duff:1986hr} and \cite{Castellani:1991et}.}
or, more recently, to test the AdS/CFT correspondence for $M_2$-branes at conical singularities \cite{Klebanov:1998hh,Dall'Agata:1999hh,Fabbri:1999hw}. Specifically, the eigenvalues of the Hodge-de Rham Laplacian,  $\Delta_2$,  can be read-off from the masses of $Z$-vector fields that arise from the Kaluza-Klein reduction of the three-form potential along  two-form harmonics.

By examining the mass spectra of $Z$-vector fields, we conclude that on each homogeneous \SE\ manifold there are two-forms with the  eigenvalues of $\Delta_2$ within the instability range \eqref{PWrange}. One must then determine whether any of those forms are basic, transverse and primitive. We found that,  given the \KK\ data for the two-form harmonics, which include the representation and the $R$-charge, it is actually the easiest to construct those forms explicitly and then verify that they indeed satisfy all the required properties. Our results are  summarized in Table \ref{tabone}, which shows that there are unstable modes for the \PW\ solution on all  homogeneous \SE\ manifolds. 

The three harmonics for the scalar fields in \eqref{rFRmass}-\eqref{PWmass} are related to the master $(1,1)$-form by operations (contractions and exterior products) that involve canonical  objects of the \SE\ geometry:  the metric and the forms, which can be expressed in terms of  Killing spinors on the  \SE\ manifold. From  a general analysis of harmonics on coset spaces with Killing spinors  \cite{D'Auria:1984vy}, it is reasonable to expect that,  at the supersymmetric solution, the three scalar fields and the $Z$-vector field should lie in the same $\cals N=2$ supermultiplet. Indeed, the pattern of masses  in \eqref{rFRmass}, the presence of the $Z$-vector field with the correct mass, and the  $R$-charges values of the four fields  suggest that it is a long $Z$-vector supermultiplet \cite{Ceresole:1984hr}. Ultimately, this observation explains why we can diagonalize the mass operator for fluctuations around the \PW\ solution on such a small set of modes -- the mixing due to the background flux involves only harmonics within a single supermultiplet. It also suggests where to look for an instability of the \PW\ solution on a general \SE\ manifold.

A regular  \SE\ manifold, $M_7$, is a $\rm U(1)$ fibration over its Kahler-Einstein base, $B_6$, so any (1,1)-form, $\omega$, as above is  a pull-back of a transverse, primitive (1,1)-form on $B_6$ with the same eigenvalue of the corresponding Hodge-de Rham Laplacian, $\Delta_{(1,1)}$. This shows that the potential instability of the \PW\ solution that we have identified, on a regular \SE\ manifold  resides in the spectrum of $\Delta_{(1,1)}$ on the Kahler-Einstein base. It also provides a link to a different class of solutions whose stability has been analyzed recently. It turns out that precisely same type (1,1)-forms, albeit with a different ``window of instability,'' were shown in  \cite{Martin:2008pf} to destabilize the $AdS_5\times B_6$ solutions \cite{Dolan:1984hv,Pope:1988xj}  of eleven-dimensional supergravity. 

Two  results in \cite{Martin:2008pf}  are directly applicable to our analysis. The first one is an explicit construction of a $(1,1)$-form $\omega$, with $\lambda_\omega=16$, on $S^2\times S^2\times S^2$, which is the \KE\ base for $Q^{1,1,1}$.  The second one is more general and concerns the spectrum of $\Delta_{(1,1)}$ on a product of two Kahler manifolds, $B_6=B_2\times B_4$. It is shown that if  $B_4$ admits a continous symmetry, then there exists a transverse, primitive $(1,1)$-form $\omega$ on $B_6$ with the eigenvalue $\lambda_\omega=16$. In particular, the unstable modes on $M^{3,2}$, which is a $\rm U(1)$ fibration over $S^2\times \CC\PP^2$,  arise  in this way. Another  Kahler-Einstein manifold that is covered by this construction is  $S^2\times dP_3$, where $dP_3$ is the del Pezzo surface. This gives us an example of  an inhomogeneous \SE\ manifold with an unstable \PW\ solution.

The rest of the paper is organized as follows. In section~\ref{sectwo}, we review the \FR\ and \PW\ solutions together with some pertinent \SE\ geometry. The details    of our calculation leading to 
the mass formulae \eqref{rFRmass}-\eqref{PWmass} are presented in section~\ref{secthr}.  In section~\ref{secexam}, we construct explicitly the unstable modes for all homogeneous examples. We conclude with some comments in section~\ref{seccom}. Our conventions and some useful identities are summarized in appendices.

\section{The solutions}
\setcounter{equation}{0}
\label{sectwo}

The \FR\  and  \PW\   solutions of eleven-dimensional supergravity on a \SE\ manifold, $M_7$, can be derived from the following general Ansatz,
\begin{equation}\label{metranz}
ds_{11}^2\eql ds_{AdS_4(L)}^2+a^2\,ds_7^2\,,
\end{equation}
\begin{equation}\label{PWflux}
\cals F_{(4)}\eql f_0\,{\rm vol}_{AdS_4(L)}+ f_i\,\Phi_{(4)} \,. 
\end{equation}
A regular \SE\ manifold, $M_7$, is the total space of a $\rm U(1)$ fibration over a  Kahler-Einstein (\KE) base, $B_6$, and the internal metric in \eqref{metranz} can  be written locally as 
\begin{equation}\label{intmetr}
ds_7^2\eql ds_{B_6}^2+c^2\big(d\psi+ A\,\big)^2\,,
\end{equation}
were $ A$ is the Kahler potential on $B_6$, $\psi$ is the angle along the  fiber and  $c$ is the squashing parameter. The potential for the internal flux in \eqref{PWflux} is given by the real part of a canonical complex three-form, $\Omega$, on $M_7$, such that 
\begin{equation}\label{intflux}
\Phi_{(4)}\eql d(\Omega+\overline \Omega)\,.
\end{equation}
The  constants, $a$, $c$, $f_0$ and $f_i$ in \eqref{metranz}-\eqref{PWflux} are  fixed by the equations of motion in terms of the $AdS_4$ radius, $L$, which  sets the overall scale of the solution:
\begin{itemize}
\item Supersymmetric and skew-whiffed \FR\ solutions
\begin{equation}\label{FRsols}
a\eql 2L\,,\qquad c\eql 1\,,\qquad f_0\eql \kappa\, {3\over 2L}\,,\qquad f_i\eql 0\,,
\end{equation}
where $\kappa=-1$ and $+1$, respectively.
\item \PW\ solution
\begin{equation}\label{PWsols}
a\eql 2\sqrt{2\over 3}\,L\,,\qquad c\eql \sqrt 2\,,\qquad f_0\eql {\sqrt 3\over 2 L}\,,\qquad f_i\eql 
{4\over 3}\sqrt{2\over 3}\,L^3\,.
\end{equation}
\end{itemize}
 
In \eqref{metranz}, we have factored out  the overall scale, $a^2$,  of the internal metric so the \KE\ metric, $g_{B_6}$, and the \SE\ metric, $\go_{M_7}$, obtained by setting $c=1$ in \eqref{intmetr}, are canonically normalized with
\begin{equation}\label{}
{Ric}(g_{B_6})\eql 8\,g_{B_6}\,,\qquad Ric(\go_{M_7})\eql 6\,\go_{M_7}\,.
\end{equation}
In the following, we will refer to the \SE\ metric on  $M_7$ as the ``round'' metric.

The one form $\vartheta=d\psi+A$, called the contact form, is globally defined on $M_7$, and is dual to the Reeb vector field, $\xi=\partial_\psi$, which is nowhere vanishing and has length one. The other two globally defined  forms of the \SE\ geometry are  the real two form, $J$, and a complex three form, $\Omega$, with its complex conjugate, $\overline \Omega$. They satisfy
\begin{equation}\label{diffid}
d\vartheta\eql 2J\,,\qquad d\Omega\eql 4 i \,\vartheta \wedge \Omega\,.
\end{equation}
Note that the ansatz \eqref{metranz}-\eqref{intflux} is in fact written in terms of globally defined objects of the  \SE\ geometry.  

It is convenient to choose   special frames, $\eo^a$, $a=1,\ldots,7$, on $M_7$, that are orthonormal with respect to the round metric and such that 
\begin{equation}\label{SEobjects}
\begin{split}
 J   \eql {i\over 2}\,\left(\eo^{z_1}\wedge \eo^{\bar z_1}+\eo^{z_2}\wedge \eo^{\bar z_2}+\eo^{z_3}\wedge \eo^{\bar z_3}\right)\,,\qquad 
 \Omega  \eql e^{4i\psi}\, \eo^{z_1}\wedge\eo^{z_2}\wedge\eo^{z_3}\,,\qquad \vartheta \eql\eo^7\,,
\end{split}
\end{equation}
where
\begin{equation}\label{cmplxfr}
\eo^{z_1}\eql \eo^1+i\,\eo^2\,,\qquad
\eo^{z_2}\eql \eo^3+i\,\eo^4\,,\qquad
\eo^{z_3}\eql \eo^5+i\,\eo^6\,,
\end{equation}
is a local holomorphic frame on the \KE\ base. This shows that  $J$ is the pull-back of the Kahler form, while $\Omega$ is,  up to a phase along the fiber,  the pull-back of the holomorphic $(3,0)$-form on $B_6$. The components of the round metric and of the squashed metric \eqref{intmetr} are   $\go_{ab}\eql\delta_{ab}$ and   $g_{ab}$, respectively.  Then the components of the eleven-dimensional metric \eqref{metranz} along the internal manifold are $\frak g_{ab}=a^2\,g_{ab}$.

One can also express $\vartheta$, $J$ and $\Omega$ as bilinears in   Killing spinors, $\eta^\alpha$,
\begin{equation}\label{Kspeqs}
\Do_a\eta^\alpha\eql {i\over 2}\Gamma_a\eta^\alpha\,, \qquad \bar\eta^\alpha\eta^\beta\eql \delta^{\alpha\beta}\,,\qquad \alpha,\beta=1,2\,,
\end{equation}
that are globally defined on $M_7$ and whose existence is  equivalent to  $M_7$ being a \SE\ manifold. In terms of $\eta^\alpha$'s we have (see, e.g., \cite{Kim:2006ag})
\begin{equation}\label{etJKsp}
\vartheta_a\eql i\,\bar\eta^1\Gamma_a\eta^2\,,\qquad J_{ab}\eql \bar\eta^1\Gamma_{ab}\eta^2\,,
\qquad \Omega_{abc}\eql -{1\over 2}(\bar\eta^1+i\,\bar\eta^2)\Gamma_{abc}(\eta^1+i\,\eta^2)\,.
\end{equation}
Using this realization together with Fierz identities, it is straightforward to prove a number of useful identities   summarized in appendix \ref{appendixB}.

To verify the solutions \eqref{FRsols} and \eqref{PWsols}, we  note that the covariant derivatives for the squashed and round metrics are related by 
\begin{equation}\label{covder}
D_aV_b\eql \Do_a V_b-2\,(c^2-1) \vartheta_{(a} J_{b)}{}^c\,V_c\,,
\end{equation}
where we have adopted a convention to raise and lower indices with the round metric, $\go_{ab}$.
For the Ricci tensors,  using identities in appendix \ref{appendixB}, we have
\begin{equation}\label{}
R_{ab}\eql\Ro_{ab}+2(1-c^2)\go_{ab}+2(3c^4+c^2-4)\vartheta_a\vartheta_b\,.
\end{equation}
These are also the components of the Ricci tensor, $\cals R_{ab}$, along the internal manifold. The Ricci tensor for $AdS_4$ of radius, $L$,  is
\begin{equation}\label{}
Ric_{AdS_4}\eql -{3\over L^2}g_{AdS_4}\,.
\end{equation}
so that the eleven-dimensional Ricci scalar is
\begin{equation}\label{Ricsc}
\cals R\eql -{12\over L^2}+{6\over a^2}(8-c^2)\,.
\end{equation}

The energy momentum tensor in \eqref{Mth:eineqs} has only diagonal contributions from the flux along $AdS_4$ and $M_7$  that are straightforward to evaluate. Then the Einstein equations \eqref{Mth:eineqs} reduce  to three algebraic equations:
\begin{equation}\label{einalg}
\begin{split}
a^2   \eql {4\over 3}(c^2-4)L^2\,,\qquad   f_0^2 & \eql {3(7c^2-16)\over 4L^2(c^2-4)}\,,\qquad f_i^2\eql {2\over 27}c^2(c^2-4)^3(c^2-1)L^6\,,
\end{split}
\end{equation}
for the size of the internal part of the metric and the parameters of the flux.

We now turn to the Maxwell equations \eqref{Mth:maxeqs}. Let us denote by  
$\ast$ the Hodge dual  
on $M_7$ with respect to the  round metric with the volume form  \begin{equation}\label{volM}
{\rm vol}_{M_7}\eql  {1\over 6}\,J\wedge J\wedge J\wedge\vartheta\eql {3\over 8}\,i\,\Omega\wedge\overline\Omega\wedge\vartheta\,.
\end{equation}
The volume form for the  squashed metric is then $c\,{\rm vol}_{M_7}$, while $ca^7{\rm vol}_{ AdS_4}\wedge{\rm vol}_{M_7}$ is the volume form in eleven-dimensions.

It follows from \eqref{diffid} and \eqref{volM} that
\begin{equation}\label{starOmega}
\ast d\Omega\eql  4 \,\Omega\,,\qquad \ast\, \Omega\eql {1\over 4}\,d\Omega\,.
\end{equation}
Then for the flux, $\cals F_{(4)}$, in \eqref{PWflux} and \eqref{intflux}, we have 
\begin{equation}\label{}
\star\cals F_{(4)}\eql f_0\star{\rm vol}_{ AdS_4}-{4f_i\over ac}\,{\rm vol}_{AdS_4}\wedge(\Omega+\overline\Omega)\,,
\end{equation}
so that
\begin{equation}\label{}
d\star\cals F_{(4)}\eql -{4f_i\over ac}\,{\rm vol}_{ AdS_4}\wedge \Phi_{(4)}\,.
\end{equation}
The second term in \eqref{Mth:maxeqs} yields
\begin{equation}\label{}
\cals F_{(4)}\wedge \cals  F_{(4)}\eql 2 f_0f_i\,{\rm vol}_{ AdS_4}\wedge \Phi_{(4)}\,,
\end{equation}
which shows that  the Maxwell equations reduce to a single equation
\begin{equation}\label{algmax}
f_i\,\Big(f_0-{2\over ac}\Big)\eql 0\,.
\end{equation}

Assuming that both $a$ and $c$ are positive, one verifies that  \eqref{FRsols} and \eqref{PWsols} exhaust all solutions to  \eqref{einalg} and \eqref{algmax}. Note that the only difference between the  supersymmetric and skew-whiffed \FR\ solutions is the sign, $\kappa$, of the flux along $AdS_4$. Equivalently, one could  reverse the orientation of the internal manifold, which changes the sign of the Hodge dual in \eqref{Mth:maxeqs}. Here, we will keep the orientation of $M_7$ fixed as in 
\eqref{volM}.

\section{The linearized analysis}
\setcounter{equation}{0}
\label{secthr}

We will not attempt here a complete analysis of the Kaluza-Klein spectrum around the \PW\ solution, but  instead will   identify a small set  of harmonics for the low lying scalar modes on which the   scalar mass operator in the linearized expansions around both the \FR\ and \PW\ backgrounds can be diagonalized. In doing that, we will be guided both by the explicit structure of the linearized equations of motion and by the properties of unstable modes on $S^7$ that were identified in  \cite{Bobev:2010ib}.

The  scalar modes we want to consider correspond  to  fluctuations of the  internal metric and  the internal three-form potential,
\begin{equation}\label{fluctuate}
  \delta \frak g_{ab}\eql \varphi (x)h _{ab}\,,\qquad \delta \cals A_{(3)}\eql  \varphi (x)\, \alpha_{(3)} \,,
\end{equation}
where $\varphi(x)$ is a scalar field  on $AdS_4$, while $h_{ab}$ and $\alpha_{(3)}$ are, respectively,  a symmetric tensor and a three form harmonic  on $M_7$. 

\subsection{Linearized Einstein equations}
\label{EinEqsLin}

We begin with the metric harmonic and the linearization of the Einstein equations \eqref{Mth:eineqs}.
Following a crucial obervation in  \cite{Bobev:2010ib} for the unstable modes on $S^7$, we will assume that 
metric harmonic, $h_{ab}$, corresponds to  a deformation of the internal metric along the $(2,0)$ and $(0,2)$ components on the \KE\ base. Specifically,    $h_{ab}$ is horizontal, that is  $\xi^ah_{ab}=0$,  and its only nonvanishing components in the basis \eqref{cmplxfr} are  $h_{z_iz_j} $ and $h_{\bar z_i\bar z_j}$. It is then automaticaly traceless. Finally, we will assume that it is transverse with respect to the round metric, $\Do^ah_{ab}=0$.  It follows then from  \eqref{covder} that it is  also transverse with respect to the internal metric with any value of the squashing parameter, $c$. 

With those assumptions, the   metric fluctuation \eqref{fluctuate} is both transverse and traceless in eleven dimensions, so that the expansion of the Ricci tensor in \eqref{Mth:eineqs} yields only one term with the Lichnerowicz operator (see, e.g., \cite{Gibbons:1978ji}), and there are no terms from  the Ricci scalar to linear order. The eleven-dimensional Lichnerowicz operator becomes then a sum,\footnote{We use $\Box_{AdS_4}=g^{\mu\nu}\nabla_\mu\nabla_\mu$, but $\Delta_L\eql-g^{ab}D_aD_b+\ldots$} 
\begin{equation}\label{}
\Box_{AdS_4}-{1\over a^2}\Delta_L^c\,,
\end{equation}
where $\Delta_L^c$ is the Lichnerowicz operator on $M_7$ with respect to the squashed metric. Then on the metric harmonics, $h_{ab}$, as above,
\begin{equation}\label{}
\Delta_L^c h_{ab}\eql \left[\Delta_L +4(1-c^2)+\Big(1-{1\over c^2}\Big)\cals L_\xi^2\right]h_{ab}\,,
\end{equation}
where $\Delta_L$ is the Lichnerowicz operator for the round metric, and $\cals L_\xi$ is the Lie derivative along the Reeb vector.

Let's denote the combination on the left hand side in \eqref{Mth:eineqs} by $\cals E_{MN}$. After  collecting all the terms in the expansion and using \eqref{Ricsc}, we obtain
\begin{equation}\label{eineqsten}
\delta\cals E_{ab}\eql -{1\over 2}\Big(\Box_{AdS_4}-\nu^2\Big)\,\varphi(x)\,h_{ab}\,,
\end{equation}
where
\begin{equation}\label{einexp}
\nu^2\eql {1\over a^2}\Delta_L+\Big({84\over a^2}-{24\over L^2}\Big)-{1\over a^2}\Big(1-{1\over c^2}\Big)(16c^2-\cals L_\xi^2)\,.
\end{equation}

Let us now turn to the expansion of the energy  momentum tensor, $\cals T_{MN}$, on the right hand side in \eqref{Mth:eineqs}. For the metric variation as above, the only terms that contribute to the linear expansion of the energy momentum tensor come  from the flux, 
\begin{equation}\label{emexp}
\delta\cals T_{ab}\eql {f_i\over 3a^6}\,g^{cf}g^{dg}g^{dh}(d\alpha_{acde}\Phi_b{}_{fgh}+d\alpha_{bcde}\Phi_a{}_{fgh})\varphi(x)  \,.
\end{equation}
Assuming that metric harmonic is an eigentensor of the Lichnerowicz operator in \eqref{einexp}, we see that
in order to diagonalize the linearized Einstein equations we must find a flux harmonic such that the symmetric tensor  in \eqref{emexp} is proportional to  $h_{ab}$. 

\subsection{Linearized  Maxwell equations}

The expansion of the Maxwell equations is
\begin{equation}\label{MaxLin}
d\star\delta\cals F_{(4)}+2\,\cals F_{(4)}\wedge\delta \cals F_{(4)}+d\,\delta(\star)\cals F_{(4)}\eql 0\,,
\end{equation}
where
\begin{equation}\label{}
\delta\cals F_{(4)}\eql d\varphi \wedge \alpha_{(3)}+\varphi \,d\alpha_{(3)}\,,
\end{equation}
and $\delta(\star)$ is the variation of the Hodge dual due to fluctuation of the metric. Define a four form
\begin{equation}\label{dotoper}
(\delta \frak g\cdot \cals F)_{MNPQ}\eql \frak g^{M'M''}\delta\frak g_{MM'}\cals F_{M''NPQ}+\ldots+\frak g^{Q'Q''} \delta\frak g_{QQ'}\cals F_{MNPQ''}\,.
\end{equation}
Then for a traceless fluctuation of the metric,
\begin{equation}\label{}
\delta(\star)\cals F_{(4)}\eql -\star(\delta\frak g\cdot \cals F_{(4)})\,.
\end{equation}

Specializing to the background flux \eqref{PWflux} and the fluctuations \eqref{fluctuate}, the linearization \eqref{MaxLin} splits into terms that are one, three, and four forms along $AdS_4$, respectively. They yield the following equations
\begin{equation}\label{otmxeqs}
d\ast_c\alpha\eql 0\,,\qquad \alpha\wedge \Phi_{(4)}\eql 0\,,
\end{equation}
and 
\begin{equation}\label{linmaxwell}
(\Box_{AdS_4}\varphi)\ast_c\alpha_{(3)}+\varphi\left[{2f_0\over a}d\alpha_{(3)}-{1\over a^2}\,d\ast_cd\alpha_{(3)}+{f_i\over a^4}\,d\ast_c(h\cdot \Phi_{(4)})\right]\eql 0\,,
\end{equation}
where $\ast_c$ denotes the dual with respect to the squashed metric and $h\cdot\Phi_{(4)}$ is defined as in 
\eqref{dotoper} using the round metric.  The various factors of the internal radius, $a$, in \eqref{linmaxwell}
are consistent with  the overall $1/L^2$  dependence of the mass terms on the $AdS_4$ radius.

\subsection{The master harmonic}
\label{masterhar}

Our task now is to identify the smallest set of harmonics on which we can diagonalize the Maxwell equation \eqref{linmaxwell}. The first step will be to streamline the evaluation of the Hodge duals.

Any $k$-form, $\Xi_k$, on $M_7$  can be uniquely decomposed into the sum, 
\begin{equation}\label{}
\Xi_k\eql \omega_k+\vartheta\wedge\omega_{k-1}\,, 
\end{equation}
 where $\omega_k$ and $\omega_{k-1}$ are horizontal forms, that is, $\imath_\xi\omega_k= \imath_\xi\omega_{k-1}=0$. Then 
\begin{equation}\label{starc}
\ast_c\Xi_k\eql c\,\ast\omega_k+{1\over c}\ast(\vartheta\wedge\omega_{k-1})\,,
\end{equation}
where $\ast$ is the Hodge dual with respect to the round metric. We may simplify this further  by introducing another Hodge dual, $\bullet$, in the space perpendicular to the fiber, or, equivalently on the \KE\ base, $B_6$.   Then for a horizontal form, $\omega$, using ${\rm vol}_{M_7}\eql {\rm vol}_{B_6}\wedge \vartheta$, we have\footnote{Note that on a $k$-form, $\ast^2=\ast_c^2=1$, while $\bullet^2=(-1)^k$.} 
\begin{equation}\label{dualss}
\ast\omega\eql \vartheta\wedge \bullet\,\omega \,,\qquad 
\ast\, (\omega\wedge\vartheta )\eql \bullet\,\omega\,,
\end{equation}
and hence
\begin{equation}\label{starbull}
\ast_c\Xi_k\eql c\,\vartheta\wedge\bullet\, \omega_k
+{1\over c}\,(-1)^{k-1}\bullet\,\omega_{k-1}\,.\end{equation}

To further restrict the Ansatz for the flux harmonic, let us look at the last term in \eqref{linmaxwell}, which is already constrained by the conditions we have imposed in section \ref{EinEqsLin} on the metric harmonic, $h_{ab}$. Since $h_{ab}$ has nonvanishing components only along the \KE\ base, we have
\begin{equation}\label{fourform}
h\cdot  \Phi_{(4)}\eql 4i\,\vartheta\wedge h\cdot(\Omega-\overline\Omega)\,,
\end{equation}
where all contractions between the metric harmonic and the background flux form are with the round metric. 

We can now evaluate the forms \eqref{fourform}  on $S^7$ using the metric harmonics given in \cite{Bobev:2010ib}. It turns out that 
$h\cdot(\Omega-\overline\Omega)$ is    closed (!)  and   it is both horizontal and invariant along the fiber. This means that it is a closed basic three-form with   nonvanishing $(2,1)$ and $(1,2)$ components.  On $S^7$, it is then a pull-back of the corresponding  closed form on $\CC\PP^3$ and thus is exact. Indeed, we find that
\begin{equation}\label{htoomega}
h\cdot(\Omega-\overline\Omega)\eql -64\, d\omega\,,
\end{equation}
where $\omega$ is a basic, primitive, transverse (1,1)-form, and an eigenform of the Laplacian, with the eigenvalue 24.  
In the following we will show that a similar construction can be carried out on a general \SE\ manifold, $M_7$. 

We start with a primitive, $(1,1)$-form, $\omega$ on the \KE\ base which is a  transverse  eigenform of the Hodge-de Rham Laplacian with the eigenvalue $\lambda_\omega$.  Its pull-back to $M_7$ is then a basic form, satisfying 
\begin{equation}\label{basicst}
\imath _\xi\omega\eql0\,,\qquad \cals L_\xi\omega\eql 0\,,
\end{equation}
which we  also denote by  $\omega$. We will now discuss the conditions on $\omega$ and derive some identities that are used later.
\smallskip

\noindent (i) The condition that $\omega$  is a primitive  $(1,1)$-form means that
\begin{equation}\label{primi}
J^{ab}\omega_{ab}\eql 0\,,\qquad J_a{}^cJ_b{}^d\omega_{cd}\eql \omega_{cd}\,,
\end{equation}
where the first equation can be equivalently written as 
\begin{equation}\label{}
J\wedge \bullet\,\omega\eql 0\qquad \text{or}\qquad J\wedge J\wedge\omega\eql 0\,.
\end{equation}
It follows from \eqref{primi} that on $B_6$ and $M_7$, respectively,\footnote{The operator $\stsx J\wedge$ on a six-dimensional Kahler manifold maps two-forms into two-forms. It has eigenvalues $-1$, $1$ and $2$ with degeneracies $8$, $6$ and $1$, respectively, corresponding to the primitive $(1,1)$-forms, $(2,0)+(0,2)$-forms and $(1,1)$-forms proportional to $J$.} 
\begin{equation}\label{primitive}
J\wedge\omega \eql-\stsx \omega\qquad \text{and}\qquad \stsv (J\wedge \omega)\eql -\vartheta\wedge\omega\,.
\end{equation}

\noindent 
(ii) Transversality on the \KE\ base
\begin{equation}\label{transsx}
d\stsx\omega\eql 0\,,
\end{equation}
implies transversality on the \SE\ manifold,
\begin{equation}\label{}
\begin{split}
d\stsv\omega & \eql d(\vartheta\wedge\stsx\omega)\\
&\eql 2J\wedge \stsx\omega-\vartheta\wedge d\stsx\omega\\
& \eql 0\,,
\end{split}
\end{equation}
where the last step follows from \eqref{primitive} and \eqref{transsx}.

By taking the exterior derivative of \eqref{primitive}, we get
\begin{equation}\label{}
J\wedge d\omega\eql 0\,,
\end{equation}
and a somewhat less obvious
\begin{equation}\label{lessobv}
J\wedge\stsx d\omega\eql 0\,.
\end{equation}
 Since the last identity is on the  \KE\ base, upon taking a dual we obtain a 1-form with components proportional to
\begin{equation}\label{jstdomx}
\begin{split}
2 J^{\alpha\beta}\nabla_{ \alpha}\omega_{\beta\gamma }+
J^{\alpha\beta}\nabla_{ \gamma}\omega_{\alpha\beta}\,.
\end{split}
\end{equation}
On a K\"ahler manifold,  $J$ is covariantly constant  and the first term can be written as
\begin{equation}\label{}
\begin{split}
J^{\alpha\beta}\nabla_{ \alpha}\omega_{\beta\gamma } & \eql \nabla^\alpha(J_ \alpha{}^{\beta}\omega_{\beta\gamma})\\
& \eql -\nabla^\alpha(J_\gamma{}^\beta\omega_{\alpha\beta})\\
& \eql 0\,,
\end{split}
\end{equation}
where we used that $\omega$ is a transverse $(1,1)$-form. The vanishing of the second term in \eqref{jstdomx} is shown similarly.
\smallskip

\noindent
(iii) Finally,  $\omega$ is an eigenfunction of the Hodge-de Rham Laplacian, $\Delta_{(1,1)}$ on $B_6$, which for a transverse form is simply, 
\begin{equation}\label{deloneone}
\Delta_{(1,1)}\omega\equiv \stsx d\stsx d\omega\eql \lambda_\omega\,\omega\,.
\end{equation}
Then the  Laplacian on  $M_7$, after using \eqref{dualss} and \eqref{lessobv} is
\begin{equation}\label{}
\begin{split}
\Delta_2 \,\omega & \equiv -\stsv d\stsv d\omega \\ & \eql -\stsv d (\vartheta\wedge \stsx d\omega)\\
& \eql  -\stsv (2J\wedge \stsx d\omega-\vartheta\wedge d\stsx d\omega)\\
& \eql \lambda_\omega\,\omega\,.
\end{split}
\end{equation}
Hence $\omega$ is also an eigenfunction of the Laplacian on $M_7$ with the same eigenvalue, $\lambda_\omega$.

\subsection{The metric harmonic}

We now take the following  Ansatz for the metric harmonic   in terms of a pure imaginary  $(1,1)$-form, $\omega$,
\begin{equation}\label{hfromdomega}
h_{ab}\eql (d\omega)_{acd}(\Omega_b{}^{cd}-\overline\Omega_b{}^{cd})+(a\leftrightarrow b)\,.
\end{equation}
This tensor is manifestly horizontal and has only $(2,0)$ and $(0,2)$ components as we have required in section 
\ref{EinEqsLin}. 
 It also satisfies  \eqref{htoomega}, as one can verify using  identities in section \ref{masterhar} and appendix~\ref{appendixA}. We will now show that $h_{ab}$ is a transverse eigentensor of the Lichnerowicz operator on $M_7$,
\begin{equation}\label{lapandshift}
\Delta _L h_{ab}\eql \lambda_h h_{ab}\,,\qquad \lambda_h\eql \lambda_\omega+4\,,
\end{equation}
with the eigenvalue, $\lambda_h$,  fixed by $\lambda_\omega$.

Before we present a somewhat lengthy proof, let us note that the same relation between the eigenvalues  of the Hodge-de Rham Laplacian and the Lichnerowicz operator has been derived  in \cite{D'Auria:1984vy} through a general analysis of the fermion/boson mass relations on manifolds with Killing spinors. In particular,  it was shown   that if a two-form and a symmetric tensor harmonics arise  from the same spin-3/2 harmonic by a supersymmetry transformation generated by   Killing spinors, the resulting shift of the eigenvalues is precisely the one given in \eqref{lapandshift}. While we have not derived  the intermediate  spin-3/2 harmonic in general,   some explicit checks on $S^7$ (or, more generally on tri-Saskian manifolds), where all the forms in \eqref{hfromdomega} can be realized in terms of Killing spinors,\footnote{See, section~\ref{trisasaki}.}  have convinced us that our construction here and in the following sections yields a subset of harmonics in a single $\cals N=2$ supermultiplet as in \cite{D'Auria:1984vy}. We will discuss it further in section~\ref{secexam}, where we   identify this supermultiplet as the long $Z$-vector multiplet \cite{Ceresole:1984hr}.

We also note that a similar construction   for  tensor harmonics on a five-dimensional \SE\ manifolds has been recently carried out in  \cite{Eager:2012hx} and  it follows a much earlier  construction for   four-dimensional Kahler manifolds  in \cite{Pope:1982ad}.

\subsubsection{Proof of transversality}

There are four types of terms in the transversality condition,\footnote{Throughout this section, $D_a$ is the covariant derivative with respect to the round metric.} 
$D^ah_{ab}=0$.  First, we have
\begin{equation}\label{}
D^a(d\omega)_{acd}\Omega_b{}^{cd}\eql-\lambda_\omega \omega_{cd}\Omega_b{}^{cd}\eql 0\,,
\end{equation}
since $\omega$ is a $(1,1)$-form. 
Secondly, 
\begin{equation}\label{}
(d\omega)_{acd}D^a\Omega_b{}^{cd}\eql 4i\,(d\omega)^{acd}\vartheta_{[a}\Omega_{bcd]}\eql0\,,
\end{equation}
since $d\omega$ is horizontal, and hence $d\omega^{abc}\vartheta_a=0$. Similarly, the full contraction between $d\omega$, which is a sum  of a $(2,1)$ and a $(1,2)$ form, and the $(3,0)$ form, $\Omega$, must vanish.
The third type of terms are
\begin{equation}\label{}
D^a(d\omega)_{bcd}\Omega_a{}^{cd}  \eql D_a(d\omega)_{bcd}\Omega^{acd}\,.
\end{equation}
Since
$D_{[a}(d\omega)_{bcd]}= 0$, 
we have
\begin{equation}\label{}
\begin{split}
3D_a(d\omega)_{bcd}\Omega^{acd} & \eql D_b (d\omega)_{acd}\Omega^{acd}\\
& \eql D_b\left[ (d\omega)_{acd}\Omega^{acd}\right]-(d\omega)_{acd}D_b\Omega^{acd}\\
& \eql -4i\,(d\omega)^{acd}\vartheta_{[b}\Omega_{acd]}\\
& \eql 0\,,
\end{split}
\end{equation}
as $d\omega$ is either contracted with $\vartheta$ or fully contracted with $\Omega$.
Finally, the last type of terms are
\begin{equation}\label{}
(d\omega)_{bcd}D^a\Omega_{a}{}^{cd}\eql 0\,,
\end{equation}
since $\Omega$ is itself transverse, see, e.g.,  \eqref{derid}.
Transversality of the terms with $\overline\Omega$ is verified similarly.

\subsubsection{Proof of \eqref{lapandshift}}

The Lichnerowicz operator,\footnote{For a list of properties of the Lichnerowicz operator, see, e.g., \cite{Gibbons:1978ji}.} $\Delta_L$, on $k$-forms coincides with the Hodge-de Rham Laplacian,
\begin{equation}\label{}
\Delta\eql d\delta+\delta d\,,\qquad \delta\eql (-1)^k *d*\,.
\end{equation}
We have assumed that $
\Delta\omega\eql\lambda_\omega\omega$.  
Using \eqref{starOmega} we also find
\begin{equation}\label{}
\Delta\Omega\eql 16\,\Omega\,.
\end{equation}

For an arbitrary tensor, the Lichnerowicz operator is defined by  
\begin{equation}\label{Lichdef}
\Delta_L T_{a_1\ldots a_k}\eql -\Box T_{a_1\ldots a_k}+(R^a{}_{a_1}T_{a\,a_2\ldots a_k}+\ldots)
-2 ( R^a{}_{a_1}{}^b{}_{a_2}T_{a\,b\,a_3\ldots a_k}+\ldots\,)\,,\end{equation}
where there are $k$-terms in the first bracket and ${1\over 2}k(k-1)$ in the second. 
An important property, which we are going to exploit in the following, is  that $\Delta_L$ commutes with the contraction.

Consider the tensor
\begin{equation}\label{}
t_{acdbef}\eql (d\omega)_{acd}\Omega_{bef}\,,
\end{equation}
from which the $(2,0)$-part of $h_{ab}$ is obtained by contracting over the pairs $ce$ and $df$ and then symmetrizing over $ab$. It follows from the definition \eqref{Lichdef} that
\begin{equation}\label{Licht}
(\Delta_Lt)_{acdbef}\eql (\Delta_L d\omega)_{acd}\Omega_{bef}+(d\omega)_{acd}(\Delta_L\Omega)_{bef}-2 D^g(d\omega)_{acd}D_g\Omega_{bef}+R\text{-terms}\,,\end{equation}
where the $R$-terms involve split contractions with both $d\omega$ and $\Omega$,
\begin{equation}\label{}
R\text{-terms}\eql -2\left[(d\omega)_{gcd}\Omega_{hef}R^g{}_a{}^h{}_b+8\text{-terms}\right]
\end{equation}
We will now show that all terms in \eqref{Licht} give contributions to $\Delta_Lh_{ab}$ that are  proportional to $h_{ab}$ and evaluate the proportionality constants. 

From the first two terms we  get $(\lambda_\omega+16)h_{ab}$.
Next, we consider the  $R$-terms, which can be traded for covariant derivatives acting on $\Omega$ using
\begin{equation}\label{}
[D_a,D_b]\Omega_{cde}\eql -\Omega_{fde}R^f{}_{cab}-\ldots\,.
\end{equation}
This gives
\begin{equation}\label{}
R\text{-terms}\eql 2\big((d\omega)_{gcd}[D^g,D_a]+(d\omega)_{agd}[D^g,D_c]+(d\omega)_{acg}[D^g,D_d]\big)\Omega_{bef}\,.
\end{equation}
The covariant derivatives acting on $\Omega$ can be evaluated using  \eqref{derid}. 
This yields terms that are products of the form
\begin{equation}\label{typetrms}
d\omega_{\times\times\times} J_{\times\times}\Omega_{\times\times\times}\qquad\text{or}\qquad
d\omega_{\times\times\times} \vartheta_\times\vartheta_\times\Omega_{\times\times\times}\,.
\end{equation}
Performing the contractions as in  the definition of $h_{ab}$, see \eqref{hfromdomega}, we are left with two free indices with all other ones contracted. Because of the symmetrization,   the free indices in the terms of the first type in \eqref{typetrms} must be on two different tensors. In particular, this implies that $J$ is always contracted with either $\Omega$ or $d\omega$ or both. All terms in which $J$ is contracted with $\Omega$ are simplified using \eqref{contrid}
and yield terms proportional to $h_{ab}$. This leaves terms in which $J$ is contracted with $d\omega$. By inspection, in all those terms $d\omega$ is doubly contracted with $\Omega$, which means that the contraction with $J$ is  once more a multiplication by $i$. 
The second type terms in \eqref{typetrms} all vanish except when the two $\vartheta$'s are contracted. Collecting all the terms we find that the total contribution  from the $R$-terms to $\Delta_Lh_{ab}$ is $-10h_{ab}$.

Finally, we consider the third term in \eqref{Licht}. Since  $d\omega$ is closed, we   rewrite this term as 
\begin{equation}\label{diffterm}
-2D_g(d\omega)_{acd}D_g\Omega_{b}{}^{cd}+(a\leftrightarrow b)\eql -2D_a(d\omega)_{gcd}D_g\Omega_{b}{}^{cd}-4D_c(d\omega)_{agd}D^g\Omega_{b}{}^{cd}+(a\leftrightarrow b)\,.
\end{equation}
Let's start with the first term in  \eqref{diffterm}. Since
\begin{equation}\label{}
(d\omega)_{gcd}D^g\Omega_{b}{}^{cd}\eql 4i\,(d\omega)^{g}{}_{cd}\vartheta_{[g}\Omega_{bcd]}\eql 0\,,
\end{equation}
we have
\begin{equation}\label{}
D_a(d\omega)_{gcd}D^g\Omega_{b}{}^{cd}\eql -(d\omega)_{gcd}D_aD^g\Omega_{b}{}^{cd}\,.
\end{equation}
Expanding the covariant derivatives using \eqref{derid}, we find that all terms involving $\vartheta$ vanish. The remaining terms have $d\omega$ contracted with $J$ and twice contracted with $\Omega$, which reduces the contraction with $J$ to the multiplication by $i$. Then the net contribution from this term to $\Delta_Lh_{ab}$ is $-6h_{ab}$.

This leaves us with  the second term in \eqref{diffterm}, which we once more rewrite using the Leibnitz rule. However, now the total derivative term does not vanish, but yields the derivative $D^c$ of the following terms,
\begin{equation}\label{expeqs}
(d\omega)_{agd}D^g\Omega_{bc}{}^{d}\eql i  (d\omega)_{agd}(2\,\vartheta^g\Omega_{bc}{}^{d}+
\vartheta_b\Omega_c{}^{gd}-\vartheta_c\Omega_{b}{}^{gd})\,.
\end{equation}
The first term on the rhs vanishes as $d\omega$ is horizontal. The second term can be rewritten as
\begin{equation}\label{}
\begin{split}
i(d\omega)_{agd}\vartheta_b\Omega_c{}^{gd} & \eql i \,h_{ac}\vartheta_b-i(d\omega)_{cgd}\Omega_{a}^{gd}\vartheta_b\,.
\end{split}
\end{equation}
Acting with $D^c$ and using 
\begin{equation}\label{omeig}
3\,D^aD_{[a}\omega_{bc]}\eql D^a(d\omega)_{abc}\eql -\lambda_\omega\,\omega_{bc}\,,
\end{equation}
and the transversality of $h_{ab}$,  we get
\begin{equation}\label{}
ih_{ac}J_{c}{}^{b}-i(d\omega)_{cgd}D_c\Omega_{a}{}^{gd}\vartheta_b-i(d\omega)_{cgd}\Omega_{a}{}^{gd}J^{c}{}_{b}\eql h_{ab}+0-(d\omega)_{bgd}\Omega_{a}{}^{gd}\,,
\end{equation}
which gives $-4h_{ab}$ contribution in $\Delta_Lh_{ab}$. The last term in \eqref{expeqs} is
\begin{equation}\label{thefgg}
-iD^c(\vartheta_c(d\omega)_{agd}\Omega_b{}^{gd})\eql 
-i \vartheta_c\,D^c(d\omega)_{agd}\Omega_b{}^{gd}-i \vartheta_c (d\omega)_{agd} D^c\Omega_b{}^{gd} \,.
\end{equation}
Using $d^2\omega=0$, the first term on the right hand side above can be simplified using
\begin{equation}\label{}
\begin{split}
-i\vartheta_c D^c(d\omega)_{agd} & \eql -i\vartheta_cD_a(d\omega)^c{}_{gd}-i\vartheta_c D_g(d\omega)_a{}^c{}_d-i\vartheta_cD_d(d\omega)_{ag}{}^c\\
&\eql iJ_{ac}(d\omega)^c{}_{gd}+iJ_{gc}(d\omega)_a{}^c{}_d+i J_{dc}(d\omega)_{ag}{}^c\\
& \eql (d\omega)_{agd}\,,
\end{split}
\end{equation}
where the second line follows using the Leibnitz rule, horizontality of $d\omega$ and \eqref{derid}. 
The second   term in \eqref{thefgg}, using \eqref{derid}, is
\begin{equation}\label{}
-i \vartheta_c (d\omega)_{agd} D^c\Omega_b{}^{gd}\eql (d\omega)_{agd}\Omega_b{}^{gd}\,.
\end{equation}
Hence the last term in \eqref{expeqs} is $-2(d\omega)_{agd}\Omega_b{}^{gd}$, and by \eqref{diffterm} it contributes $-8h_{ab}$ to $\Delta_Lh_{ab}$.

Finally, using the Leibnitz rule, we are left with
\begin{equation}\label{}
4(d\omega)_{agd}D_cD_g\Omega_{bcd}\eql 16\,h_{ab}\,.
\end{equation}
Hence all terms in \eqref{Licht} are indeed proportional to $h_{ab}$, with the net result
\begin{equation}\label{}
\lambda_h\eql\lambda_\omega+16-10-6-4-8+16\eql\lambda_\omega+4\,.
\end{equation}
This concludes the proof of \eqref{lapandshift}.

\subsection{The flux harmonics}

We take as internal flux harmonic the linear combination 
\begin{equation}\label{fluxanz}
\alpha_{(3)}\eql t_1\,\vartheta\wedge \omega +t_2\, \ast d\,(\vartheta\wedge \omega )\,,
\end{equation}
where $t_1$ and $t_2$ are arbitrary pure imaginary  parameters.\footnote{In the following, we   denote this harmonic   by $\alpha$.}

The harmonics that arise in the expansion of the Maxwell equation 
\eqref{linmaxwell}  are: $d\alpha$ , $\ast_c\alpha$, and $d\ast_c d\alpha$. We will now show that for $\alpha$ given by \eqref{fluxanz}, each of those terms is a linear combination of the following two linearly independent harmonics: 
\begin{equation}\label{basfms}
\Lambda_1=\stsv(\vartheta\wedge \omega)\,\qquad\text{and}\qquad \Lambda_2=d(\vartheta\wedge\omega)\,.
\end{equation}
Specifically, we find
\begin{align}
d\alpha & \eql \lambda_\omega\, t_2 \, \Lambda_1+(t_1-2t_2) \,\Lambda_2\,,\label{LAone}\\[6 pt]
\ast_c\alpha & \eql {1\over c}\left( {t_1 } +2\,  t_2 \,(c^2-1) \right) \Lambda_1+  c\,t_2\,\Lambda_2\,,\label{LAtwo}\\[6 pt]
d\ast_c d\alpha & \eql    {\lambda_\omega\over  c}(t _1-2t_2)\,\Lambda_1 
+  c \left(\lambda_\omega t_2-2(t_1-2t_2) \right) \Lambda_2
\,.\label{LAthr}
\end{align}

The first identity follows from
\begin{equation}\label{}
\begin{split}
d\ast d(\vartheta\wedge\omega) & \eql d\ast(2J\wedge \omega)-d\ast(\vartheta\wedge d\omega)\\
& \eql -2\,d(\vartheta\wedge\omega)+d\stsx d\omega\\
& \eql -2 \,d(\vartheta\wedge\omega)+\lambda_\omega \stsx \omega\\
& \eql -2 \,d(\vartheta\wedge\omega)+\lambda_\omega \stsv (\vartheta\wedge\omega)\,.
\end{split}
\end{equation}
where we used \eqref{diffid}, \eqref{dualss}, \eqref{primitive} and \eqref{deloneone}. The second one is an immediate consequence of \eqref{starc}, \eqref{diffid} and \eqref{primitive}. For the third one, we have
\begin{equation}\label{}
\begin{split}
d\ast_cd\,\alpha & \eql t_2\lambda_\omega d\,\ast_c \stsv (\vartheta\wedge\omega)+(t_1-2t_2) \,d\ast_cd(\vartheta\wedge\omega)\\
& \eql  c\,\lambda_\omega\,t_2 \,d(\vartheta\wedge\omega)+(t_1-2t_2){1\over c}\left[d\stsv d(\vartheta\wedge \omega)+2(c^2-1)\,d(\vartheta\wedge\omega)\right]\\
& \eql {\lambda_\omega\over c}(t_1-2t_2)\stsv (\vartheta\wedge\omega) 
- c\,\left[2(t_1-2t_2)-\lambda_\omega t_2\right]\,d(\vartheta\wedge\omega)
\,.
\end{split}
\end{equation}

In evaluating the contribution from the metric fluctuation to the linearized Maxwell equations \eqref{linmaxwell} we also need the indentity 
\begin{equation}\label{hPhiid}
h\cdot \Phi_{(4)}\eql -128\,i\,\vartheta\wedge d\omega\,.
\end{equation}
To prove it, we note that by the second identity in \eqref{diffid},
\begin{equation}\label{Phiexp}
\Phi_{(4)}\eql  4i\vartheta\wedge (\Omega-\overline\Omega)\,.
\end{equation}
Since $h_{ab}$ is horizontal,
\begin{equation}\label{}
h\cdot\Phi_{(4)} \eql -i(h\cdot\Omega)\wedge\vartheta+i(h\cdot\overline \Omega)\wedge\vartheta\,,
\end{equation}
where $(h\cdot\Omega)_{abc}\eql 3 h_{d[a}\Omega_{bc]}{}^d$. Using the definition  \eqref{hfromdomega} and the identity \eqref{bomom}, we find that only $\overline\Omega$ terms in $h_{ab}$ contribute  to the contraction $h\cdot\Omega$. The three terms in that contraction  are then evaluated using  \eqref{bbarom} and \eqref{bigomid}. The result is given in   \eqref{htoomega}, but now we have shown that it holds on any \SE\ manifold. Including the conjugate terms yields  \eqref{hPhiid}.

Finally,
\begin{equation}\label{dasas}
d\ast_c(\vartheta\wedge d\omega)\eql - {\lambda_\omega\over c}\stsv(\vartheta\wedge\omega)\eql 
 - {\lambda_\omega\over c}\,\Lambda_1\,.
\end{equation}
This proves that all terms in \eqref{linmaxwell}  are  linear combinations of the two basis harmonics \eqref{basfms}.

It also follows from \eqref{dasas} that $d\Lambda_1=0$. Since $d\Lambda_2=0$ as well, we have $d\ast_c\alpha=0$ as required by \eqref{otmxeqs}. The other equation in \eqref{otmxeqs} is satisfied automatically.

We must also evaluate the linearized   energy momentum tensor \eqref{emexp}. To this end we note that the two basis harmonics \eqref{basfms}, using  \eqref{diffid} and \eqref{primitive}, can be  written as 
\begin{equation}\label{}
\Lambda_1\eql -J\wedge \omega\,,\qquad \Lambda_2\eql -2\Lambda_1-\vartheta\wedge d\omega\,.
\end{equation}
Hence $d\alpha$ in \eqref{LAone} is a linear combination of a horizontal $(2,2)$-form  $J\wedge\omega$ and a mixed form  $\vartheta\wedge d\omega$. Given \eqref{Phiexp}, the contraction  in \eqref{emexp} with   $J\wedge\omega$ must vanish. Similarly, the only nonvanishing terms in the contraction with the second form are those in which the free indices are along the base and the two $\vartheta$'s are contracted. This gives
\begin{equation}\label{}
g^{cf}g^{dg}g^{eh}(\vartheta\wedge\omega)_{acde}\Phi_{fgh}\eql {12\, i\over c^2}(d\omega)_{ade}(\Omega b{}^{de}-\overline\Omega_b{}^{de})\,,
\end{equation}
where the indices on the right hand side are raised with the round metric.  The full expansion of \eqref{emexp} is then
\begin{equation}\label{linem}
\delta\cals T_{ab}\eql -{4\,i\over a^6c^2}\,f_i\,(t_1-2t_2)\varphi(x)\,h_{ab}\,,
\end{equation}
and is indeed proportional to the metric harmonic.

\subsection{The masses}

For a scalar field, $\varphi(x)$,  satisfying  \eqref{sceqs} with mass, $m$, and the metric and flux harmonics as above, the linearized Einstein equations \eqref{eineqsten}-\eqref{emexp}  become diagonal, 
\begin{equation}\label{algein}
-{1\over 2}\left[{m^2}-{1\over a^2} (\lambda_\omega+4) +{24\over L^2}+{4\over a^2}\left(4c^2-{4\over c^2}-21\right)\right]h_{ab}\eql -4i\,{f_i\over a^6\,c^2}(t_1-2\, t_2)\,h_{ab}\,.
\end{equation}
To evaluate the left hand side, we have used  \eqref{lapandshift} and $\cals L_\xi^2h_{ab}=-16 h_{ab}$. The latter follows from the observation that the $R$-charge of the metric harmonic is $q=4$ and is the same as of the background flux. The contraction in the fluctuation of the energy momentum tensor on the right hand side  has been evaluated in \eqref{linem}.

The linearized Maxwell equation \eqref{linmaxwell} can be simplified using \eqref{LAone}-\eqref{LAthr}. After projecting onto the  basis harmonics, $\Lambda_1$ and $\Lambda_2$, it yields two equations
\begin{equation}\label{algmaxeqs}
\begin{split}
{1\over  c}\Big({m^2}-{\lambda_\omega\over a^2}\Big)\,t_1
+\left[{2\over  c}(c^2-1){m^2}+2\,\lambda_\omega\,\Big({1\over   a^2c}+{f_0\over a}\Big)\right]\,t_2 & \eql -128\,i\,{f_i\over a^4}\,{\lambda_\omega\over  c}\,,
\\[6 pt]
2\,\Big({  c\over a^2}+{f_0\over a}\Big) \,t_1+\left[  c\,\Big({m^2}-{\lambda_\omega\over a^2}-{4\over a^2}\Big)-4\,{f_0\over a}\right]\,t_2 & \eql 0\,.
\end{split}
\end{equation}

For the \FR\ solutions \eqref{FRsols} there is no internal flux, $f_i=0$, and the Einstein and Maxwell equations decouple. From the first one we get the same mass,  
\begin{equation}\label{FRmassg}
m_1^2L^2\eql {\lambda_\omega\over 4}-2 ,
\end{equation}
 for both the supersymmetric and skew-whiffed solution. The other two masses in \eqref{rFRmass} and \eqref{swFRmass}  are then obtained by setting the determinant of  the homogeneous system of equations \eqref{algmaxeqs} for $t_1$ and $t_2$ to zero. This yields a quadratic equation for  $m^2$, whose solutions are either 
 \begin{equation}\label{FRmassfl}
m_2^2L^2\eql{\lambda_\omega\over 4}+\sqrt{\lambda_\omega+1}-1\,,\qquad m_3^2L^2\eql{\lambda_\omega\over 4}-\sqrt{\lambda_\omega+1}-1\,,
\end{equation}
for the supersymmetric or
\begin{equation}\label{}
m_2^2L^2\eql{\lambda_\omega\over 4}+2\sqrt{\lambda_\omega+1}+2\,,\qquad 
m_3^2L^2\eql {\lambda_\omega\over 4}-2\sqrt{\lambda_\omega+1}+2\,,
\end{equation}
for the skew-whiffed solutions, respectively. 

For the \PW\ solution, all three equations are coupled by the non-vanishing internal flux.
Solving \eqref{algmaxeqs} for $t_1$ and $t_2$ and plaguing into \eqref{algein} yields a cubic equation for $m^2$, whose solutions are
\begin{equation}\label{themasses}
m_1^2L^2\eql {3\over 8}\lambda_\omega\,,\qquad 
m_2^2L^2\eql {3\over 8}\lambda_\omega  +3\sqrt{1+\lambda_\omega}+3\,,\qquad
m_3^2L^2\eql {3\over 8}\lambda_\omega  -3\sqrt{1+\lambda_\omega}+3\,.
\end{equation}
For each of the masses there is a fluctuation of the metric and the flux that together diagonalize  the linearized equations of motion around the  \PW\ solution.
As we have already discussed in section \ref{Intro}, the last mass will violate the BF bound when $\lambda_\omega$ lies in the range \eqref{PWrange}.  
One may note that the masses $m_2^2$ and $m_3^2$ for the \PW\ solution are simply 3/2 of the masses for the flux modes in the skew-whiffed \FR\ solution.

\section{Examples}
\setcounter{equation}{0}
\label{secexam}

In this section we will construct explicitly the  master form(s), $\omega$, leading to an instability of the \PW\ solutions for two classes of \SE\ manifolds: the tri-Sasakian manifolds and the  homogeneous manifolds \eqref{homSE}.  Throughout this section we take $\omega$ to be real. The unstable modes in section~\ref{secthr} are then constructed using the form $i\,\omega$.

\subsection{Tri-Sasakian manifolds}
\label{trisasaki}

The eleven-dimensional supergravity admits a consistent truncation on an arbitrary tri-Sasakian manifold to a 
$\cals N=3$, $d=4$ gauged supergravity  \cite{Cassani:2011fu}. As shown in \cite{Cassani:2011fu}, the instability of the \PW\ solution follows then from the existence of a single scalar mode with the mass $m^2=-3$ in the spectrum of fluctuations around the corresponding critical point of the scalar potential. 

Starting with that unstable scalar mode in the four-dimensional theory, one can  follow the truncation and reconstruct the unstable mode in eleven-dimensions. However, it is simpler to look directly for a  $(1,1)$-form, $\omega$, in terms of the geometric data on a tri-Sasakian manifold.

A tri-Sasakian manifold  admits three globally defined orthonormal Killing spinors, $\eta^i$, in terms of which  the three  one-forms, $K^i$, dual to the $\rm SU(2)$ Killing vectors, are given by
\begin{equation}\label{Killthsas}
K^i_a\eql {i\over 2}\epsilon^{ijk}\,\bar\eta^j\Gamma_a\eta^k\,.
\end{equation}
Define 
\begin{equation}\label{}
M^{i}\eql -{1\over 2}dK^i\,,\qquad  M^i_{ab}\eql -{1\over 2} \epsilon^{ijk}\,\bar\eta^j\Gamma_{ab}\eta^k\,.
\end{equation}
The forms $K^i$ and $M^i$ satisfy   \cite{Pope:1985bu}
\begin{align}
\label{idDK}
\Do_aK_b^i & \eql -M_{ab}^i\,,\\ 
\label{idDM} 
\Do_aM_{bc}^i & \eql 2\go_{a[b}K_{c]}^i\,,\\
\label{idKK}
K_a^iK^{ja} & \eql\delta^{ij}\,,\\
\label{idMK}
M^i_{ab}K^{jb}& \eql\epsilon^{ijk}K^k_a\,,\\
\label{idMM}
M^i_{ac}M^{jc}{}_b & \eql K^i_aK^j_b-\delta^{ij}\go_{ab}+\epsilon^{ijk}M^k_{ab}\,.
\end{align}
Using those identities we show that the two-forms
\begin{equation}\label{MKforms}
\omega^i\eql {1\over 2}\,\epsilon^{ijk}\,K^j\wedge K^k+{1\over 3}\,M^i\,,
\end{equation}
are transverse eigenforms of the Hodge-de Rham Laplacian,
\begin{equation}\label{lapontS}
\Delta_2\, \omega^i\eql 24\,\omega^i\,.
\end{equation}

Indeed, the transversality follows directly from \eqref{idDK}-\eqref{idMK}, which imply that
\begin{equation}\label{}
\Do^aM^i_{ab}\eql 6 K^i_b\,,\qquad \Do^a(K^j_{[a}K^{k\phantom{j}}_{b]})\eql -\epsilon^{jki}K^i_{b}\,.
\end{equation}
To prove \eqref{lapontS}, we note that on a transverse form, $\omega^i$, 
\begin{equation}\label{laponome}
\Delta\omega^i_{ab}\eql -\Do^c(d\omega^i)_{abc}\,,
\end{equation}
where 
\begin{equation}\label{}
d\omega^i\eql -2 \epsilon^{ijk}M^j\wedge K^k\,.
\end{equation}
The  divergence in \eqref{laponome} is then evaluated by first using \eqref{idDK} and \eqref{idDM} and then simplifying the resulting contractions using \eqref{idKK}-\eqref{idMM}.

The \PW\ solution is now obtained by choosing any two orthonormal Killing spinors that fix a particular \SE\ structure. Given the $\rm SU(2)$ isometry, we may simply take $(\eta^\alpha)=(\eta^1,\eta^2)$ and set $\chi=\eta^3$ to be the additional Killing spinor. Then $\vartheta=K^3$ and $J\eql -M^3$, see \eqref{etJKsp}. Consider  the two form
\begin{equation}\label{trisasom}
\omega\eql  K^1\wedge K^2- {1\over 3}\, J\,, 
\end{equation}
with components
\begin{equation}\label{}
\omega_{ab}\eql -2  (\bar\eta^1\Gamma_{[a}\chi)(\bar\eta^2\Gamma_{b]}\chi)-  {1\over 3}\bar\eta^1\Gamma_{ab}\eta^2\,.
\end{equation}
It follows from \eqref{idKK} and \eqref{idMK} that $\omega$ is horizontal. Similarly, the form  
\begin{equation}\label{}
d(K^1\wedge K^2)=-2(M^1\wedge K^2-K^1\wedge M^2)\,,
\end{equation}
is horizontal, so that $\omega$ is in fact basic. Finally, by contracting with $J$, we check that $\omega$ is a primitive (1,1)-form. 

We have checked that the unstable mode  arising from  $\omega$ in \eqref{trisasom}   reproduces precisely the unstable mode  in the truncation in \cite{Cassani:2011fu}. We also note that a more complete construction and classification of harmonics on $N^{1,1}$ in terms of Killing spinors, including the forms above,  can be found in  \cite{Billo:2000zs}. 

While the construction above gives an unstable mode on any tri-Sasakian manifold, there will be additional modes if the manifold admits more than three Killing spinors.\footnote{In fact, the only regular manifold with more than three Kiling spinors is $S^7$ \cite{Boyer:2007nr}.} In particular, to construct the unstable modes on $S^7$ found in \cite{Bobev:2010ib}, we can generalize the foregoing  as follows. Let
 $\chi^j$ be the additional six Killing spinors and let
\begin{equation}\label{}
K^{\alpha j}_a\eql i\,\bar \eta^\alpha\Gamma_a\chi^j\,,\qquad
\alpha=1,2\,,\quad j=1,\ldots,6\,.
\end{equation}
Then
\begin{equation}\label{}
\omega^{ij}\eql {1\over 2}( K^{1i}\wedge   K^{2j}+ K^{1j}\wedge K^{2i})- {1\over 3} J\,\delta^{ij}\,,
\end{equation}
are symmetric, $\omega^{ij}=\omega^{ji}$,  and traceless, $\omega^{ij}\delta_{ij}=0$, and transform in ${\bf 20'}$ of $\rm SU(4)$, which is the isometry of the \KE\ base, $\CC\PP^3$. In the same way as above, one checks that $\omega^{ij}$ are basic (1,1)-forms and that the diagonal forms, $\omega^{jj}$, are transverse and $\Delta_2\,\omega^{jj}\eql 24\,\omega^{jj}$. By the $\rm SU(4)$ symmetry, the same   holds for the remaining forms.  

\subsection{Homogeneous Sasaki-Einstein manifolds}

The homogeneous \SE\ manifolds  \eqref{homSE} are given by  $G/H$ coset spaces, which is a convenient realization for a calculation of the \KK\ spectrum of the corresponding  $AdS_4\times M_7$ compactification of  eleven dimensional supergravity. However,  one can also realize any homogeneous \SE\ manifold as a hypersurface in some $\CC^N$, in some cases modded out by a continuous Abelian symmetry. This has been discussed in detail in \cite{Klebanov:1998hh,Dall'Agata:1999hh,Fabbri:1999hw,Fre':1999xp,Ceresole:1999zg,Gauntlett:2005jb,Klebanov:2010tj}. In this section we use the latter construction to  find explicitly   stability violating master  $(1,1)$-forms, $\omega$, on each of the spaces \eqref{homSE}.  An advantage of this method is that the required properties of $\omega$ are either manifest or  easy to verify.

\begin{table}[t]
\begin{center}
{
\begin{tabular}{@{\extracolsep{20 pt}} c c c c c c}
\toprule
\noalign{\smallskip}
Spin & Field & Energy & ${\rm U(1)}_R$ & $m^2L^2$ \\
\noalign{\smallskip}
\midrule
\noalign{\smallskip}
1	& $Z$ & $E_0+1$ & $q$ & $4 E_0(E_0-1)$ \\
0 & $\pi$ & $E_0+2$ & $q$ & $(E_0+2)(E_0-1)$\\
0 & $\phi$ & $E_0+1$ & $q+4$  & $(E_0+1)(E_0-2)$ \\
0 & $\phi$ & $E_0+1$ & $q$  & $(E_0+1)(E_0-2)$ \\
0 & $\phi$ & $E_0+1$ & $q-4$  & $(E_0+1)(E_0-2)$ \\
0 & $\pi$ & $E_0$ & $q$ & $E_0(E_0-3)$\\
\bottomrule
\end{tabular}
}
\caption{\label{tabtwo} The bosonic sector of a $Z$-vector multiplet.}
\end{center}
\end{table}

In principle, one could try to identify  $(1,1)$-forms leading to  instabilities of   \PW\ solutions by examining  the  \KK\ spectra that have been studied for all homogeneous \SE\ manifolds in references in Table \ref{tabone}. Indeed, in the \KK\ reduction of the three-form potential, $\cals A_{(3)}$, a transverse two-form harmonic gives rise to a vector field whose mass is given by the eigenvalue of the Hodge-de Rham Laplacian \cite{Castellani:1984vv,Duff:1984sv}. In the terminology of \cite{D'Auria:1984vy}, the vector field is called the $Z$-vector field and it  is  present in the \KK\ towers  of  the following $\cals N=2$ supermultiplets \cite{Ceresole:1984hr,Fabbri:1999mk}: the long and/or semi-long graviton multiplet, the two long and/or semi-long gravitino multiplets and the $Z$-vector multiplet.

However,  the mere  presence in the \KK\ spectrum  of a  two-form harmonic, $\omega$,  whose mass, $\lambda_\omega$,  lies in the instability range \eqref{PWrange}, is  not yet sufficient to conclude that the \PW\ solution is unstable. One must also show that $\omega$ is a transverse, primitive, basic,  (1,1)-form, which is by no means obvious. For that reason, we first construct explicitly stability violating $(1,1)$-forms, $\omega$,  and then check whether both $\lambda_\omega$ and the supersymmetric \FR\ scalar masses \eqref{FRmassg} and \eqref{FRmassfl} agree with the known \KK\ spectra, in particular, whether  the corresponding fields: the $Z$-vector field, the scalar and the two pseudo-scalar fields lie in a long $Z$-vector supermultiplet. The comparison works perfectly for $S^7$, $N^{1,1}$ and $M^{3,2}$, but  reveals missing supermultiplets in the published \KK~spectra for $Q^{1,1,1}$ and $V^{5,2}$.

The bosonic fields of a long  $Z$-vector supermultiplet are listed in Table~\ref{tabtwo}, with the $R$-charge in the second column and the masses in the last column given in the conventions used in this paper. Specifically, the $R$-charge  is twice the charge in the original tables in the \KK\ literature  (see, e.g., Table 3 in  \cite{Fabbri:1999mk}). We define the mass of a $Z$-vector as the eigenvalue of the corresponding Hodge-de Rham Laplacian. This agrees with the usual definition used in the  references in Table~\ref{tabone}, except that our normalization  of the metric for the \FR\ solution introduces a factor of four difference,
\begin{equation}\label{}
m^2_Z L^2\eql {1\over 4} {M_Z^2\over e^2}\,.
\end{equation}
The masses of the scalar fields are related by 
\begin{equation}\label{}
 m^2_{\phi,\pi} L^2\eql {1\over 16}\Big({M^2_{\phi,\pi}\over e^2}-32\Big)\,,
\end{equation}
where $e^2=1/(16L^2)$ is usually set to one.

\subsubsection{$S^7$}

We represent $S^7$ as the unit sphere in $\CC^4$,
\begin{equation}\label{sphere}
|u^1|^2+\ldots +|u^4|^2\eql 1\,.
\end{equation}
The $\rm U(1)_R$ symmetry is the rotation by the phase. Let  $\Phi_{ij\bar k\bar l}$ be a constant complex tensor in $\CC^4$ that is antisymmetric in $[ij]$ and $[\bar k\bar l]$, primitive with respect to the canonical complex structure in $\CC^4$,  and satisfies the reality condition $\Phi_{ij\bar k\bar l}=(\Phi_{kl\bar i\bar j})^*$. Then the pull-back onto $S^7$ of 
\begin{equation}\label{ssevforms}
\omega\eql \Phi_{ij\bar k \bar l}\,u^i\bar u{}^{\bar k}\,du^j\wedge d\bar u {}^{\bar l}\,,
\end{equation}
yields 20  basic $(1,1)$-forms  with $\lambda_\omega=24$, which give rise to the unstable modes obtained in \cite{Bobev:2010ib}. 

Our calculation agrees with the general result for the spectrum of the Hodge-de Rham Laplacian on $\CC\PP^3$  \cite{cp3eigenvalues},  conveniently summarized in Table 2 in \cite{Martin:2008pf}. There we find that   there is a single tower of $(1,1)$-forms in $[k,2,k]$ irrep of $\rm SU(4)$ with the eigenvalues 
\begin{equation}\label{cpthreela}
\lambda_{(1,1)}\eql 4(k+2)(k+3)\,,\qquad k=0,1,2,\ldots\,.
\end{equation}
The forms  \eqref{ssevforms} lie at the bottom of the tower with $k=0$. The higher level forms with $k\geq 1$  have $\lambda_{(1,1)}\geq 48$ and thus lie outside the instability bound \eqref{PWrange}.

One may note that those forms are not the lowest lying transverse two-forms on $S^7$. Indeed, the spectrum of the Laplacian on two-forms on $S^7$ is  \cite{Duff:1986hr} 
\begin{equation}\label{}
\lambda_{(2)}\eql (p+2)(p+4)\,,\qquad p=1,2,3,\ldots\,,
\end{equation}
of which \eqref{cpthreela} is a subset. For $p=1$ and $2$, the eigenvalues are $15 $ and $24$, respectively, and satisfy \eqref{PWrange}.  However, the two-forms with $\lambda_{(2)}=15$ are trilinear in $u^i$ and $\bar u^i$, hence  have a nonzero $R$-charge and are not basic.
\subsubsection{$N^{1,1}$}

The (hyper)Kahler quotient construction for $N^{1,1}$ \cite{Fre':1999xp,Gauntlett:2005jb} starts with $\CC^3\oplus\overline \CC{}^3$ with  coordinates $ (u^j,v_j)$, $j=1,2,3$,  that transform   as  $\bf 3$ and $\bf \bar 3$ under $\rm SU(3)$, respectively, and with $(u^j,-\bar v^j)$  transforming as doublets under $\rm SU(2)$. The $N^{1,1}$ manifold is then the  surface 
\begin{equation}\label{}
|u^j|^2\eql |v_j|^2\eql 1\,,\qquad u^jv_j\eql 0\,,
\end{equation}
modded by the  $\rm U(1)$ action $(u^i,v_i)\sim (e^{i\delta}u^i,e^{-i\delta}v_i)$. The standard metric \cite{Castellani:1983tc,Page:1984ac} is obtained by a reduction from the flat metric in $\CC^6$. We refer the reader to  \cite{Gauntlett:2005jb} for a detailed discussion of the metrics and for explicit angular coordinates.

 The three Killing forms in section \ref{trisasaki} can be taken as 
\begin{equation}\label{}
\begin{split}
K^1 & \eql {1\over 2}(u^jdv_j+\bar u _jd\bar v^j)\,,\qquad K^2   \eql -{i\over 2}(u^jdv_j-\bar u _jd\bar v^j)\,, \qquad 
K^3  \eql {i\over 2}(u^jd\bar u_j+v_jd\bar v^j)\,,
\end{split}
\end{equation}
in terms of which the master form, $\omega$, is given by \eqref{trisasom}.
It is now manifest that $\omega$ is a $(1,1)$-form, which is invariant   under the $\rm U(1)$ action of the Kahler quotient, and hence a well-defined form on $N^{1,1}$. It is also a singlet of $\rm SU(3)$  and  is invariant under the ${\rm U(1)}_{R}\subset{\rm  SU(2)}$ isometry, $(u^j,v_j)\rightarrow (e^{i\psi} u^j,e^{i\psi}v_j)$, along the \SE\ fiber.    Evaluating it in angular coordinates, we verify  that it is basic and primitive.

The complete \KK\ spectrum on this space was obtained in \cite{Termonia:1999cs} (see, also \cite{Castellani:1984gf,Fre':1999xp,Billo:2000zr,Billo:2000zs}), where one finds  21 towers of two-form harmonics. Specifying to the $\bf (1,3)$ irreducible representation  of $\rm SU(3)\times SU(2)$, $M_1=M_2=0$ and $J=1$ in the notation in \cite{Termonia:1999cs}, leaves two possible eigenvalues $\lambda_{12}^{(2)}=96$ and  $\lambda_{21}^{(2)}=48$ lying in the series $E_8$ with $j=0$. The first three forms are the ones constructed above in \eqref{MKforms}, one of which is the sought after $(1,1)$-form, $\omega$, with $\lambda_\omega=24$. The remaining three are the three canonical two-forms, $M^i$, on the tri-Sasakian manifold, one of which is the complex structure, and hence is not primitive, while the other two are not basic.

In this example the $\cals N=2$  long  $Z$-vector multiplet is a part of a long $\cals N=3$ gravitino  supermultiplet, see Table 4 in \cite{Billo:2000zs}. Following \cite{D'Auria:1984vy}, all harmonics in this multiplet can be constructed in terms of the three Killing spinors on $N^{1,1}$ \cite{Billo:2000zs}.

\subsubsection{$M^{3,2}$}

The $\cals N=2$ supersymmetry of the \FR\ solution on $M^{3,2}$ was proved in \cite{Castellani:1983mf}. The complete Kaluza-Klein spectrum was  obtained in  \cite{D'Auria:1984vv} (see also \cite{Castellani:1991et}) and further analyzed more recently in \cite{Fabbri:1999mk}. The \KE\ base  of $M^{3,2}$  is  $\CC\PP^2\times \CC\PP^1$   and  the \SE\ metric  in the form \eqref{intmetr} is given by \cite{Page:1984ad,Pope:1984ig}
\begin{equation}\label{}
ds^2\eql {3\over 4} ds^2_{\CC\PP^2}+{1\over 2}ds^2_{\CC\PP^1}+(d\psi+{3\over 4}A_{\CC\PP^2}+{1\over 2}A_{\CC\PP^1})^2\,,
\end{equation}
where the $ds_{\CC\PP^k}^2$  is the Fubini-Study metric  and $A_{\CC\PP^k}$ is the Kahler potential  with   $dA_{\CC\PP^k}\eql 2 J_{\CC\PP^k}$.

The Kahler quotient construction for this \SE\ manifold \cite{Dall'Agata:1999hh,Fabbri:1999hw} starts with  $\CC^3\oplus \CC^2$ with  coordinates, $u^i$ and $v^\alpha$, in terms of which $M^{3,2}$ is   the surface,
\begin{equation}\label{}
2\,u^j\bar u_j\eql 3\,v^\alpha\bar v_\alpha\eql 1\,,
\end{equation}
modded by the $\rm U(1)$ symmetry, $(u^i,v^\alpha)\sim(e^{2i\delta}u^i,e^{-3i\delta} v_\alpha)$. Once more the ${\rm U(1)}_R$ symmetry is 
$(u^i,v^\alpha)\rightarrow  (e^{i\psi}u^i,e^{i\psi}v^\alpha)$.

The lowest lying modes of the scalar Laplacian that are invariant under ${\rm U(1)}_R$ come from isometries of the \KE\ base. In particular,  the scalar harmonics corresponding to the ${\rm SU(3)}$ Killing vectors on $\CC\PP^3$ are given by
\begin{equation}\label{mschar}
Y\eql t_{i}{}^j u^i\bar u_j\,,
\end{equation}
where  $t_{ij}$ is a constant hermitian, traceless matrix, and satisfy  \cite{Hoxha:2000jf}
\begin{equation}\label{}
\Delta Y\eql 16\,Y\,.
\end{equation}
By a general construction of \cite{Martin:2008pf}, there is a corresponding primitive $(1,1)$-form on $\CC\PP^2\times \CC\PP^1$  with the same eigenvalue of the Hodge-de Rham Laplacian. It is given by
\begin{equation}\label{mastmthtw}
\omega\eql 2 i\,\partial_B\bar\partial_B Y +16\,Y J^{(4)}-16 \,Y J^{(2)}\,,
\end{equation}
where the  \SE\ two form is $J=J^{(4)}+J^{(2)}$ with
$J^{(4)}= {3\over 4}J_{\CC\PP^2}$ and $J^{(2)}={1\over 2}J_{\CC\PP^1}$, and  $\partial_B$ and $\bar\partial_B$ are the Dolbeault operators (see, e.g.,~\cite{Sparks:2010sn}). 
The eight  unstable scalar modes  transform in $(\bfs 8,\bfs 1)$ of $\rm SU(3)\times SU(2)$ and have  the mass 
\begin{equation}\label{}
m_3^2L^2\eql 9-3\sqrt{17}\approx -3.3693\,.
\end{equation}

In the \KK\ spectrum for the supersymmetric solution, we should find a $Z$-vector supermultiplet with the masses
\begin{equation}\label{Zmassmthtw}
m_Z^2L^2\eql 16\,,\qquad m^2_\phi L^2\eql 2 \,,\qquad m^2_\pi L^2\eql 3\pm\sqrt{17}\,.
\end{equation}
Indeed, there is such a multiplet given by Eqs.\ (3.23) and (3.24) in \cite{Fabbri:1999mk}, where we must set
$M_1=M_2=1$ and $J=0$. It has 
\begin{equation}\label{engzero}
E_0\eql {1\over 2}(1+\sqrt{17})\,,
\end{equation}
which reproduces the masses \eqref{Zmassmthtw} using formulae in Table~\ref{tabtwo}.

\subsubsection{$Q^{1,1,1}$}

 Recall that $Q^{1,1,1}$ is a $\rm U(1)$ bundle over 
$\CC\PP^1\times\CC\PP^1\times\CC\PP^1$, with the metric (see, e.g., \cite{Pope:1984ig})
\begin{equation}\label{}
ds^2\eql {1\over 2}(ds_{\CC\PP^1_{(1)}}^2+ds_{\CC\PP^1_{(2)}}^2+ds_{\CC\PP^1_{(3)}}^2)
+\left[d\psi+
{1\over 2}\,(A_{\CC\PP^1_{(1)}}+ \,A_{\CC\PP^1_{(2)}}+ \,A_{\CC\PP^1_{(3)}})\right]^2\,.
\end{equation}
The Kahler quotient construction for this manifold \cite{Dall'Agata:1999hh,Fabbri:1999hw},
 has three $\CC^2$'s, with coordinates $u^\alpha$, $v^\alpha$ and $w^\alpha$, respectively, one for   each $\CC\PP^1$ factor in the \KE\ base. Then $Q^{1,1,1}$ is the surface in $\CC^6$,
\begin{equation}\label{}
u^\alpha\bar u_\alpha\eql v^\alpha\bar v_\alpha\eql w^\alpha\bar w_\alpha\eql 1\,,
\end{equation}
modded by two $\rm U(1)$ symmetries, $(u^\alpha,v^\alpha,w^\alpha)\sim
(e^{i\delta}u^\alpha, e^{i\theta}v^\alpha,e^{-i\delta-i\theta}w^{\alpha})$. In terms of the projective coordinates, $z_i$, on $\CC\PP^1_{(i)}$, and the fiber angle, $\psi$, we  have
\begin{equation}\label{}
\begin{split}
u^1\eql {z_1 e^{2i\psi/3}\over(1-|z_1|)^{1/2}}\,,\qquad v^1\eql {z_2 e^{2i\psi/3}\over(1-|z_2|)^{1/2}}\,,\qquad w^1\eql {z_3 e^{2i\psi/3}\over(1-|z_3|)^{1/2}}\,,\\
u^2\eql { e^{2i\psi/3}\over(1-|z_1|)^{1/2}}\,,\qquad v^2\eql { e^{2i\psi/3}\over(1-|z_2|)^{1/2}}\,,\qquad w^2\eql { e^{2i\psi/3}\over(1-|z_3|)^{1/2}}\,.
\end{split}
\end{equation}

The $\rm SU(2)$ Killing vectors on each $\CC\PP^1$ yield  triplets of scalar harmonics,
\begin{equation}\label{sclharq}
Y_{(1)}\eql t^{(1)}{}_\alpha{}^\beta u^\alpha\bar  u_\beta\,,\qquad
Y_{(2)}\eql t^{(2)}{}_\alpha{}^\beta v^\alpha\bar  v_\beta\,,\qquad
Y_{(3)}\eql t^{(3)}{}_\alpha{}^\beta w^\alpha\bar  w_\beta\,,
\end{equation}
which are eigenfunctions of the Laplacian with the eigenvalue $16$ \cite{Pope:1984ig}.
The two forms
\begin{equation}\label{omqqq}
\omega_{(1)}\eql Y_{(1)} (J_{\CC\PP^1_{(2)}}-J_{\CC\PP^1_{(3)}} )\,,\qquad
\omega_{(2)}\eql Y_{(2)} (J_{\CC\PP^1_{(3)}}-J_{\CC\PP^1_{(1)}} )\,,\qquad
\omega_{(3)}\eql Y_{(3)} (J_{\CC\PP^1_{(1)}}-J_{\CC\PP^1_{(2)}} )\,, 
\end{equation}
 are primitive, transverse   $(1,1)$-eigenforms of the Hodge-de Rham Laplacian with the same eigenvalue \cite{Martin:2008pf}. This gives nine unstable modes for the \PW\ solution on $Q^{1,1,1}$   
 in the adjoint representation of $\rm SU(2)\times SU(2)\times SU(2)$.

Clearly, numerical values of all the masses of the $Z$-vector field and the scalar and pseudoscalar fields at the \FR\ solution are the same as for $M^{3,2}$, and one expects to find a similar structure of $\cals N=2$ supermultiplets as well.  Hence it is surprising that  the \KK\ spectrum in section 4 in \cite{Merlatti:2000ed} does not include  a  long $Z$-vector supermultiplet in the adjoint of $\rm SU(2)\times SU(2)\times SU(2)$ with the energy \eqref{engzero}. In fact, there is also no graviton multiplet corresponding to the scalar harmonics \eqref{sclharq}.
However, a closer examination of the allowed harmonics on $Q^{1,1,1}$ and their masses, which are listed in  section 3 of the same paper, shows that the $Z$-vector supermultiplet we are looking for should have been included in the final ``complete classification.''

\subsubsection{$V^{5,2}$}
\label{Vfivetwo}

As discussed in \cite{Klebanov:1998hh} (see, also \cite{Ceresole:1999zg,Bergman:2001qi,Martelli:2009ga}), the  Stiefel manifold, $V^{5,2}$, is the intersection of the Kahler cone in~$\CC^5$,     
\begin{equation}\label{}
(u^1)^2+(u^2)^2+(u^3)^2+(u^4)^2+(u^5)^2\eql 0\,,
\end{equation}
with the unit sphere,
\begin{equation}\label{}
|u^1|^2+|u^2|^2+|u^3|^2+|u^4|^2+|u^5|^2\eql 1\,.
\end{equation}
Writting $u^j=x^j+iy^j$, the real and imaginary part vectors  $(x^j)$ and $(y^j)$ in $\RR^5$ can be parametrized by the Euler angles of the coset space $\rm SO(5)/SO(3)$,\footnote{K.P.\ would like to thank N.\ Bobev and N.\ Warner for a discussion that led to these formulae. A somewhat different explicit parametrization of $V^{5,2}$ is given in \cite{Bergman:2001qi,Martelli:2009ga}.} 
\begin{equation}\label{coorsub}
\left(\begin{matrix}
x^1 & y^1 \\
x^2 & y^2 \\
x^3 & y^3 \\
x^4 & y^4 \\
x^5 & y^5 \\
\end{matrix}\right) \eql
\left(\begin{matrix}
\cR_3(\alpha_1,\alpha_2,\alpha_3) & 0 \\
0 & \cR_2(\phi)\\
\end{matrix}\right)
\left(\begin{matrix}
\cos\theta & 0 \\ 0 & \cos\mu \\ 0 & 0 \\ \sin\theta & 0\\ 0 & \sin\mu \\
\end{matrix}\right)\cR_2(\coeff 4 3\,\psi)\,,
\end{equation}
where 
\begin{equation}\label{}
0\leq \alpha_1,\alpha_3< 2\pi\,,\qquad 0\leq \alpha_2\,,\phi <\pi\,,\qquad -{\pi\over 2}\leq \mu\,,\theta<{\pi\over 2}\,,\qquad 0\leq \psi<{3\pi\over 8}\,,
\end{equation}
and $\cals R_2$ and $\cals R_3$ are rotation matrices. In terms of the coordinates on the cone and the angles, the \SE\ metric on $V^{5,2}$ is
\begin{equation}\label{vftmet}
\begin{split}
ds^2 & \eql {3\over 2}du^j d\bar u^j-{3\over 16} |u^jd\bar u^j|^2 \\[6 pt]
& \eql {3\over 8}\,\Big[d\mu^2+\cos^2\mu\,\sigma_1^2+d\theta^2+\cos^2\theta\,\sigma_2^2 \\&
\qquad\qquad  +
{1\over 2}\,\sin^2(\mu-\theta)(\sigma_3+d\phi)^2+{1\over 2}\,\sin^2(\mu+\theta)(\sigma_3-d\phi)^2\Big]\\[6 pt]
& \qquad\qquad\qquad +\Big[d\psi+{3\over 8}\cos(\mu-\theta)( \sigma_3+d\phi)+{3\over 8}\cos(\mu+\theta)( \sigma_3-d\phi)\Big]^2\,,
\end{split}
\end{equation}
where $\sigma_i$ are the $\rm SO(3)$-invariant forms, $d\sigma_i=\sigma_j\wedge\sigma_k$. The metric \eqref{vftmet} is the  canonical \SE\ form of the $\rm U(1)$ fibration over the \KE\ base, which is the Grassmannian, $Gr_{2}(\RR^{5})$.
 
The harmonics on $V^{5,2}$ are obtained by the pullback of  tensors in $\CC^5$ and decompose into representations of $\rm SO(5)\times U(1)_R$. Here    $\rm SO(5)$  acts on $u^j$ in the real vector representation, while  ${\rm U(1)}_R$ is the phase rotation, $u^j\rightarrow e^{i\psi} u^j$.  

The lowest lying scalar harmonic that is invariant under $\rm U(1)_R$ is
$\Phi^{ij}\eql u^i\bar u ^j-u^j\bar u^i$. It is an eigenfunction of the Laplacian with the eigenvalue 16
\cite{Ceresole:1999zg}. Similarly, the lowest lying $(1,1)$-forms that are not proportional to the Kahler form are
\begin{equation}\label{vomega}
\omega^i\eql \epsilon^{ijklm}u^j\bar u^k du^l d\bar u^m\,.
\end{equation}
They transform as $\bf 5$ of $\rm SO(5)$ and are invariant under ${\rm U(1)}_R$. Expanding those forms using \eqref{coorsub} confirms that they are  basic. They satisfy
\begin{equation}\label{}
\Delta_2\,\omega^i\eql {32\over 3}\,\omega^i\,,
\end{equation}
and hence  give rise to five unstable modes of the \PW\ solution with the mass
\begin{equation}\label{}
m_3^2L^2\eql 7-\sqrt{105}\approx - 3.2469\,.
\end{equation}

The masses for the supersymmetric solution are
\begin{equation}\label{}
m_Z^2L^2\eql {32\over 3}\,,\qquad m_\phi^2 L^2\eql {2\over 3}\,,\qquad m_\pi^2 L^2\eql {5\over 3}\pm\sqrt{35\over 3}\,.
\end{equation}
The $Z$-vector multiplet has then 
\begin{equation}\label{}
E_0\eql {1\over 6}(3+\sqrt{105})\,.
\end{equation}
While such a multiplet is not listed in the tables in \cite{Ceresole:1999zg}, the authors note at  the end of  section~2 that there might be an additional vector supermultiplet with this energy.\footnote{We thank A.\ Ceresole and G.\ Dall'Agata for correspondence, which clarified this point.} In appendix \ref{appendixC}, we list all bosonic harmonics on $V^{5,2}$ that transform in $\bfs 5$ of $\rm SO(5)$ and show that they decompose unambigously into  $\cals N=2$ supermultiplets including a long $Z$-vector supermultiplet in agreement with our construction.

\subsection{Orbifolds}

Homogeneous \SE\ manifolds also admit discrete symmetries such that the quotient maniofld, $M_7/\Gamma$ is still \SE. The natural question is what happens to the master (1,1)-forms in this projection and whether the \PW\ solution for the quotient \SE\ manifold is stable. We will now examine this for some examples of \SE\ discrete quotients that were considered in the literature.

For  $S^7$, it has been shown in  \cite{Bobev:2010ib} that if the discrete symmetry group $\Gamma$ is a subgroup of $\rm SU(4)$, it will  preserve some of the unstable modes. The same reasoning applies to the (1,1)-forms 
\eqref{ssevforms} and shows that  some of them will be well-defined on the quotient.

Orbifolds of $M^{3,2}$,  $Q^{1,1,1}$, and $S^{7}$,  can be obtained as limits of the $Y^{p,k}$ Sasaki-Einstein manifolds  \cite{Martelli:2008rt}.  
Specifically, when
 $2k=3p$ and $p=2r$, one has that $Y^{2r,3r}(\CC\PP^{2}) = M^{3,2} / \ZZ_{r}$, where $\ZZ_r$ is a finite subgroup of $ \rm SU(2)$ acting on  $\CC\PP^{1}$. Since the master 2-forms for $M^{3,2}$ are constructed from scalar harmonics on the $\CC\PP^{2}$, see \eqref{mschar} and \eqref{mastmthtw}, they are preserved under the orbifolding. Hence the instability persists for these orbifolds of $M^{3,2}$.  

Similarly, when $k=p$, one has   $Y^{p,p}( \CC\PP^{1} \times \CC\PP^{1}) = Q^{1,1,1}/\ZZ_{p}$, where $\ZZ_p$ is a  finite subgroup of  $ \rm SU(2)$ acting on one of the three $\CC\PP^1$'s. Each independent master (1,1)-form on $Q^{1,1,1}$, see \eqref{sclharq} and \eqref{omqqq}, is constructed from a scalar harmonic on one of the $\mathbb\CC\PP^{1}$ factors. For the $\rm SU(2)$ acting on $\CC\PP^1_{(i)}$, the forms  $\omega_{(j)}$, $j\not=i$, are invariant under $\ZZ_p$ and hence are well defined on the quotient $ Q^{1,1,1}/\ZZ_{p}$.

For $k=3p$, one has that $Y^{p,3p} = S^{7}/\ZZ_{3p}$, where $\ZZ_{3p}\subset {\rm SU(4)}$ acts by 
\begin{equation}\label{orbs7}
(u^{1},u^{2},u^{3},u^{4})\quad \longrightarrow \quad (e^{2\pi i/3p}u^{1},e^{2\pi i/3p}u^{2},e^{2\pi i/3p}u^{3},e^{-2\pi i/p}u^{4}).         
\end{equation}
The six master $(1,1)$-forms on $S^7$ that contain precisely one $u^4$ or $\bar u^4$ are not invariant under \eqref{orbs7}. This yields fourteen unstable modes on that space.

The orbifolds $V^{5,2}/\mathbb{Z}_{k}$ have been discussed in \cite{Martelli:2009ga}. The finite group  here is $\mathbb{Z}_{k} \subset {\rm U(1)}_{b}$, where ${\rm U(1)}_{b}$ is a diagonal subgroup of the $\rm SO(2)\times SO(2)$ rotation in the $(12)$ and $(34)$ planes in $\CC^5$. Clearly, the master 2-form $\omega^{5}$, see  \eqref{vomega}, is invariant under this action and yields one unstable mode on  $V^{5,2}/\mathbb{Z}_{k}$.

\section{Conclusions }
\label{seccom}

In this paper we have analyzed  a subset of scalar modes in the linearized spectrum of  eleven-dimensional supergravity around the Pope-Warner solution on an arbitrary \SE\ manifold and derived a   condition under which the solution   becomes perturbatively unstable. Specifically, we have shown that when the manifold admits a  basic, transverse, primitive $(1,1)$-form within a certain range of eigenvalues of the Hodge-de Rham Laplacian, then there are scalar modes violating the \BF\ bound. 
We have also constructed such  destabilizing $(1,1)$-forms on all homogenous \SE\ manifolds,  and on their orbifolds, and found that when viewed as harmonics for fluctuations around the supersymmetric solution, those forms give rise to a long $Z$-vector supermultiplet in the \KK\ spectrum. 

Throughout the paper we have assumed that the \SE\ manifold was regular, and the regularity was used explicitly in some of the proofs, in particular, in establishing  shifts between   eigenvalues of mass operators on various harmonics. However, since those proofs are  local, one would expect that our construction should hold for an arbitrary \SE\ manifold.  Equivalently, one could try to rephrase the stability condition in terms of spinor-vector harmonics on the \SE\ manifold which by the construction in \cite{D'Auria:1984vy}
give rise to long $Z$-vector supermultiplets.

It remains an open problem to see whether stability violating $(1,1)$-forms  exist on any \SE\ manifold. If the manifold is regular, this reduces to a problem of determining the low lying spectrum of the Hodge-de Rham Laplacian on  a six-dimensional \KE\ manifold, which  in itself is a difficult problem with rather few explicit results (see, e.g.,  \cite{Berger:2007}). 

The main motivation for recent interest in  \PW\ solutions came from the top down construction of holographic models of superconductors  in \cite{Gauntlett:2009dn,Gubser:2009gp,Gauntlett:2009bh}. The \PW\ solutions  are then dual to zero entropy states with emergent conformal invariance at $T=0$. We refer the reader to \cite{Donos:2011ut}  for further discussion of physical significance of various instabilities, including those at $T>0$.

There is also an  analogue  of the \PW\ solution  in   type IIB supergravity  \cite{Romans:1984an}, which is known to be unstable within  the $N=8$, $d=5$ supergravity  \cite{Pilch:1999,Girardello:1999bd} obtained by compactification on $S^5$. It would be interesting, and perhaps simpler,  to examine the stability of this type of solutions for the new class of five-dimensional \SE\ manifolds  \cite{Gauntlett:2004hh,Cvetic:2005ft}, for which the spectra of the scalar Laplacian were already obtained in  \cite{Kihara:2005nt,Oota:2005mr}.

\bigskip
\bigskip
\leftline{\bf Acknowledgements}
\smallskip
We would like to thank Nikolay Bobev and Scott MacDonald for  discussions. 
This work is supported in part by DOE grant DE-FG03-84ER-40168. 
\vfill\eject


\begin{appendices}

\section{Conventions}

\renewcommand{\theequation}{A.\arabic{equation}}
\setcounter{equation}{0}
\label{appendixA}

We use the same conventions as in \cite{Nemeschansky:2004yh} and \cite{Bobev:2010ib}, with the mostly plus space-time metric and the bosonic field equations of eleven-dimensional supergravity given in \eqref{Mth:eineqs} and \eqref{Mth:maxeqs}, and the gravitino supersymmetry transformations 
\begin{equation}\label{susytr}
\begin{split}
\delta\psi_M & \eql D_M\epsilon +{1\over 144}\left(\dGamma_M{}^{NPQR}-8\,\delta_M{}^N \dGamma^{PQR}\right)\cals F_{NPQR}\,\epsilon\,.
\end{split}
\end{equation}
 
On a manifold with a Minkowski signature metric, $\frak g$, we define the Hodge dual, $\star$, by    
\begin{equation}\label{}
\star\Lambda\wedge \Lambda\eql -|\Lambda|{\rm vol}_{\frak g}\,.
\end{equation}
The Hodge dual, $\ast$, for a riemannian metric, $g$, is then defined without the minus sign. 

The eleven-dimensional Dirac matrices in the $4+7$ decomposition are 
\begin{equation}\label{}
\begin{split}
\dGamma^\mu & \eql \gamma^{\mu-1}\otimes {\bf 1}\,,\qquad \mu=1,\ldots,4\,,\\
\dGamma^{a+4} & \eql -\gamma^5\otimes \Gamma^a\,,\quad a=1,\ldots,7\,,
\end{split}
\end{equation}
where
\begin{equation}\label{}
\gamma^5\eql i\gamma^0\gamma^1\gamma^2\gamma^3\,,\qquad \Gamma^7\eql i\,\Gamma^1\ldots\Gamma^6\,.
\end{equation}
Then
\begin{equation}\label{}
\dGamma^1\dGamma^2\ldots\dGamma^{11}\eql \epsilon^{12\ldots 11}\,\bfs 1\eql \bfs 1\,.
\end{equation}
We use the representation in which the four-dimensional $\gamma$-matrices are real, while the seven-dimensional $\Gamma$-matrices are pure imaginary and antisymmetric. For a real spinor, $\eta$, on the internal manifold, we then have $\bar\eta=\eta^T$.

\section{Sasaki-Einstein identities}

\renewcommand{\theequation}{B.\arabic{equation}}
\setcounter{equation}{0}
\label{appendixB}

In the local frame 
\begin{equation}\label{}
\begin{split}
e^{1,2,3} & \eql e^{r/L}dx^{0,1,2}\,,\qquad e^4\eql dr\,,\qquad  e^{a+4}   \eql 2L\,\eo^a\,,\quad a=1,\ldots,7\,,
\end{split}
\end{equation}
on  $AdS_4\times M_7$, cf.\ \eqref{SEobjects}, the unbroken supersymmetries are given by $\epsilon=\varepsilon\otimes \eta$, 
\begin{equation}\label{killspex}
\varepsilon\eql e^{r/L}\varepsilon_0\,,\qquad \gamma^{012}\varepsilon_0\eql \varepsilon_0\,,
\end{equation}
and
\begin{equation}\label{killspproj}
\eta\eql (\cos(2\psi)+\sin(2\psi)\Ga^{12})\eta_0\,,\qquad \Gamma^{12}\eta_0\eql\Gamma^{34}\eta_0\eql\Gamma^{56}\eta_0\,,
\end{equation}
where   $\varepsilon_0$ and $\eta_0$  are constant spinors.
We  choose the two independent solutions, $\eta^1$ and $\eta^2$, of  \eqref{killspproj} such that the components of the two \SE\ tensors  in \eqref{SEobjects} and \eqref{etJKsp} are the same.

Given the Reeb vector field of unit length,\footnote{All indices are raised and lowered with the \SE\ metric, $\go_{ab}$. 
}
\begin{equation}\label{}
\xi^a\xi_a\eql \vartheta_a\vartheta^a= 1\,,
\end{equation}
the projection operator 
\begin{equation}\label{}
\pi^a{}_b\eql\delta^a{}_b-\vartheta^a\vartheta_b\,,
\end{equation}
is a map onto the subspace perpendicular to the Reeb vector. Any tensor $H_{ab\ldots c}$ satisfying
\begin{equation}\label{}
\vartheta^aH_{ab\ldots c}\eql \vartheta^bH_{ab\ldots c}\eql \ldots \eql \vartheta^cH_{ab\ldots c}\eql 0\,,
\end{equation}
will be invariant under the projection, and, modulo its dependence on the fiber coordinate, $\psi$, can be thought of as a tensor on the Kahler-Einstein base. We refer to such tensors as horizontal. 

For complex horizontal tensors of rank $n$ there is a further decomposition into $(p,q)$-type tensors, where $p$ and $q$, $p+q=n$, refer  to the number of holomorphic and  anti-holomorphic indices according to the corresponding decomposition along the Kahler-Einstein base. In particular, $J_{ab}$ and $\Omega_{abc}$, are horizontal tensors of type $(1,1)$ and $(3,0)$, respectively. A contraction of $J$ with a $(p,0)$-type and $(0,p)$-type horizontal tensor is a multiplication by $+i$ and $-i$, respectively. For example,
\begin{equation}\label{contrid}
J_a{}^d\,\Omega_{bcd} \eql i\,\Omega_{abc}\,,\qquad J_a{}^d\, \overline \Omega_{bcd} \eql - i\,\overline \Omega_{abc}\,.
\end{equation}
Horizontal tensors (forms) that are in addition invariant along the Reeb vector field are called basic.

Using the explicit realization of the Sasaki-Einstein forms in terms of Killing spinors \eqref{etJKsp}, one can prove  additional identities, which we use frequently. First, we have the 
 following ``single contraction'' identities
\begin{gather}\label{bomom}
J^{ac}J_{bc} \eql   \pi^a{}_b\,,\qquad
\Omega^{abe}\Omega_{cde} \eql 0\,,\\ \Omega^{abe}\,\overline\Omega_{cde}  \eql 4\,  \pi^{[a}{}_{[c} \pi^{b]}{}_{d]}
-4 \, J^{[a}{}_{[c} J^{b]}{}_{d]}
-8i\,  \pi^{[a}{}_{[c} J^{b]}{}_{d]}\,,\label{bbarom}
\end{gather}
from which the higher contractions follow,
\begin{equation}\label{}
J_{ab}J^{ab}  \eql 6\,,\qquad \Omega^{acd}\,\overline\Omega_{bcd} \eql 8 \pi^a{}_b-8 i J^a{}_b\,,
\qquad \Omega^{abc}\overline\Omega{}_{abc} \eql 48\,.\end{equation}
We also need  the following uncontracted identity
\begin{equation}\label{bigomid}
\Omega^{abc}\,\overline\Omega_{def}\eql 
6\,\pi^{[a}{}_{[d}\pi^b{}_e\pi^{c]}{}_{f]}-18i\,
\pi{}^{[a}{}_{[d}\pi{}^b{}_e J{}^{c]}{}_{f]}-18\,
\pi{}^{[a}{}_{[d}J{}^b{}_e J{}^{c]}{}_{f]}+6i\,
J{}^{[a}{}_{[d}J{}^b{}_e J{}^{c]}{}_{f]}\,.
\end{equation}
and covariant derivatives of the Sasaki-Einstein forms that are given by
\begin{equation}\label{derid}
\Do_a\vartheta_b   \eql  \,J_{ab}\,,\qquad 
\Do_aJ_{bc}   \eql -2\,  \,\go_{a[b}\vartheta_{c]}\,,\qquad 
\Do_a\Omega_{bcd} \eql 4i\, \,\vartheta_{[a}\Omega_{bcd]}\,.
\end{equation}
Identities  \eqref{diffid} follow from \eqref{derid}  by  antisymmetrization.

\section{Some harmonics on $V^{5,2}$}

\renewcommand{\theequation}{C.\arabic{equation}}
\setcounter{equation}{0}
\label{appendixC}

The   classification of supermultiplets in the \KK\ spectrum on $V^{5,2}$ given in Tables 2-6  in \cite{Ceresole:1999zg} does not include any long  $Z$-vector supermultiplet. However, the discussion in section 2 in that paper suggests that  some vector multiplets might be  missing from the classification.
In this appendix,  we  use standard group theory methods (see, e.g., \cite{Castellani:1991et}) to list all harmonics on $V^{5,2}$ that transform   in $\bfs 5$ of $\rm SO(5)$. This allows us to determine unambigously that there must be  a long Z-vector supermultiplet in the \KK\ spectrum consistent with the explicit construction in section~\ref{Vfivetwo}.   We refer the reader to   \cite{Ceresole:1999zg} and the references therein for the group theoretic set-up of the harmonic analysis on this space.

The $V^{5,2}$ manifold is a $G/H$ coset space, 
\begin{equation}\label{}
V^{5,2}\eql  {\rm SO(5)\times U(1)\over SU(2)\times U(1)}\,,
\end{equation}
where the embeding of   $H$   in $G$ is defined  by the branching rule 
\begin{equation}\label{HinG}
\bfs 5_Q\quad\longrightarrow\quad \bfs 3_Q+\bfs 1_{Q+1}+\bfs 1_{Q-1}\,.
\end{equation}
It then follows that the embedding of $H$ into the tangent $\rm SO(7)$ group is given by 
\begin{align} \label{bransc}
\bfs 1 & \quad\longrightarrow\quad \bfs 1_0\,,\\ \label{branchvec}
\bfs 7 & \quad\longrightarrow\quad  \bfs 3_1+\bfs 3_{-1}+\bfs 1_0\,,\\
\bfs 8 & \quad\longrightarrow\quad  \bfs 3_{1/2}+\bfs 3_{-1/2}+\bfs 1_{3/2}+\bfs 1_{-3/2}\,.
\end{align}
This shows that the embedding is through the chain
\begin{equation}\label{}
\rm SU(2)\times U(1)\subset SU(3)\times U(1)\subset SU(4)\subset SO(7)\,,
\end{equation}
where $\rm SU(2)\subset SU(3)$ is the maximal embedding. The other two embeddings are regular, except that the normalization of the $\rm U(1)$ charge is half the conventional one \cite{Slansky:1981yr}.

In addition to \eqref{branchvec}, we also need the branchings of $\bf  21$, $\bf  35$ and $\bf  27$ of $\rm SO(7)$, which determine   the two-form, the three-form and the symmetric tensor 
harmonics, respectively,
\begin{equation}\label{otbranch}
\begin{split}
\bfs 2\bfs 1 & \quad \longrightarrow \quad \bfs 1_0+\bfs 3_{2}+ {\bfs 3}_{1}+\bfs 3_0+\bfs 3_{-1}+  {\bfs 3}_{-2}+\bfs 5_0 \,,\\
\bfs 3\bfs 5 &    \quad \longrightarrow \quad  \bfs 1_3+\bfs 1_{1}+\bfs 1_0 
+\bfs 1_{-1}+\bfs 1_{-3}
+\bfs 3_2 + {\bfs 3}_{1}+
\bfs 3_0+\bfs 3_{-1}+ {\bfs 3}_{-2}+\bfs 5_{1}+\bfs 5_0+\bfs 5_{-1}\,,\\
\bfs 2\bfs 7 & \quad \longrightarrow \quad \bfs 1_2+\bfs 1_0+\bfs 1_{-2}+\bfs 3_{1}+\bfs 3_0+ {\bfs 3}_{-1}+\bfs 5_{2}+\bfs 5_0+ {\bfs 5}_{-2}\,.
\end{split}
\end{equation}

We recall that each independent harmonic is completely specified by its $G\times H$ representation. It follows from \eqref{HinG} that only representations $\bfs 3_q$ and $\bfs 1_q$ in the branchings \eqref{bransc}, \eqref{branchvec} and \eqref{otbranch} give rise to harmonics in $\bfs 5_Q$ of ${\rm SO(5)\times U(1)}_R$. Specifically, each $\bfs 3_q$ yields a single harmonic, $(\bfs 5_q,\bfs 3_q)$, while each $\bfs 1_q$ yields two harmonics, $(\bfs 5_{q-1},\bfs 1_q)$ and $(\bfs 5_{q+1},\bfs 1_q)$. 

After compiling the list of all harmonics, one must identify the longitudinal ones, which do not give rise to four-dimensional fields in the \KK\ expansion. This can be done by looking at the representation labels of the harmonics. For example, there are two scalar harmonics in $(\bfs 5_{1},\bfs 1_0)$ and $(\bfs 5_{-1},\bfs 1_0)$, and four vector harmonics in $(\bfs 5_1,\bfs 3_1)$, $(\bfs 5_{-1},\bfs 3_{-1})$, $(\bfs 5_1,\bfs 1_0)$ and $(\bfs 5_{-1},\bfs 1_0)$. The last two  are in the same representations as the scalar harmonics and are longitudinal. Indeed, the scalar harmonics are the functions $z^i$ and $\bar z^i$, respectively, and the corresponding longitudinal vector harmonics are $dz^i$ and $d\bar z^i$. The remaining two transverse vector harmonics are obtained from $z^iz^jd\bar z^j$ and $\bar z^i\bar z^j dz^j$. The same procedure   is used to  count the two-form, the  three-form, and the symmetric tensor longitudinal harmonics.

\begin{table}[t]
\begin{center}
{
\begin{tabular}{@{\extracolsep{10 pt}} c | c c c c c c c c c}
\toprule
\noalign{\smallskip}
$Q=$ & $4$ & $3$ &   $2$ & $1$ & 0 & $-1$  & $-2$ & $-3$ & $-4$\\
\noalign{\smallskip}
\midrule
\noalign{\smallskip}
$h$ & & & &  $sg_+$ & & $sg_-$ & \\[6 pt]
$Z$ &  & & $sg_-$ & $sg_+$ & $Z$ & $sg_-$  &   $sg_+$ & \\[6 pt]
$A$ &  & &                   & $sg_+ $ & & $sg_-$ \\[6 pt]
$W$  &  & &                   & $W_+ $ & & $W_-$  \\[6 pt]
$\pi$ & $W_+$ &&$W_-$, $H$ & $W_+$ & $Z$, $Z$ & $W_-$  & $W_+$, $H$&&  $W_-$  \\[6 pt]
$\phi$ & & $Z$ & & $W_+$ & $Z$ & $W_-$ & & $Z$ \\[6 pt]
$\Sigma$ & & & & $W_+$ & & $W_-$ \\[6 pt]
$S$ & & & & $H$ & & $H$ \\[6 pt]
\bottomrule
\end{tabular}
}
\caption{\label{tabthr} 
{ The $\cals N=2$ supermultiplets  on $V^{5,2}$ in $\bfs 5$ of $\rm SO(5)$.
}}
\end{center}
\end{table}

Using \KK\ expansions in \cite{D'Auria:1984vy} (see  also \cite{Billo:2000zs} for a succinct summary), it is then straightforward to identify the four dimensional fields corresponding to the transverse harmonics and arrange them  into $\cals N=2$ supermultiplets, whose field content is given, e.g., in Tables  1-9 in \cite{Fabbri:1999mk}. The result is summarized in  Table~\ref{tabthr}, where the first column lists the four-dimensional fields. The remaining columns are labelled by the $U(1)$ charges of the $R$-symmetry subgroup of $G$. The $R$-charge in \cite{Fabbri:1999mk} is $y_0=2Q/3$. Each entry in those columns corresponds to a transverse harmonic in the $\bfs 5_Q$ representation of $\rm SO(5)\times U(1)_R$, with the symbol indicating the $\cals N=2$ supermultiplet that the corresponding four-dimensional field belongs to: $sg_\pm$ --  short graviton multiplets, $Z$ -- a long $Z$-vector multiplet, $W_\pm$ -- long $W$-vector multiplets, and $H$ -- a hypermultiplet.

\end{appendices}

\vfill\eject



\end{document}